\documentclass[iop]{emulateapj}
\newcommand{\fer}{{\it Fermi}}
\newcommand{\wse}{{\it WISE}}
\newcommand{\bzcat}{ROMA-BZCAT}
\usepackage{color}
\usepackage{graphicx}
\usepackage{longtable}
\usepackage{draftcopy}
\usepackage{threeparttable}

\slugcomment{version \today: fm}
\shorttitle{Association of $\gamma$-ray sources with \wse\ colors}
\shortauthors{R. D'Abrusco, F. Massaro, A. Paggi, N. Masetti, G. Tosti,
		      M. Giroletti \& H.~A. Smith}

\begin{document}
\title{Unveiling the nature of the unidentified $\gamma$-ray sources I: \\ a new method for the 
association of $\gamma$-ray blazars}
\author{R. D'Abrusco\altaffilmark{1}, F. Massaro\altaffilmark{2}, A. Paggi\altaffilmark{1}, N. Masetti\altaffilmark{3},
G. Tosti\altaffilmark{4,5}, M. Giroletti\altaffilmark{6} \& H.~A. Smith\altaffilmark{1}.}

\altaffiltext{1}{Harvard - Smithsonian Astrophysical Observatory, 60 Garden Street, Cambridge, MA 02138, USA}
\altaffiltext{2}{SLAC National Laboratory and Kavli Institute for Particle Astrophysics and Cosmology, 2575 
Sand Hill Road, Menlo Park, CA 94025, USA}
\altaffiltext{3}{INAF/IASF di Bologna, via Gobetti 101, I-4019 Bologna, Italy}
\altaffiltext{4}{Dipartimento di Fisica, Universit\`a degli Studi di Perugia, 06123 Perugia, Italy}
\altaffiltext{5}{Istituto Nazionale di Fisica Nucleare, Sezione di Perugia, 06123 Perugia, Italy}
\altaffiltext{6}{INAF Istituto di Radioastronomia, via Gobetti 101, 40129, Bologna, Italy}

\begin{abstract}

We present a new method for identifying blazar candidates by examining the \emph{locus}, 
i.e. the region occupied by the \fer\ $\gamma$-ray blazars in the three-dimensional 
color space defined by the \wse\ infrared colors.
This method is a refinement of our previous approach that made use of the 
two-dimensional projection of the distribution of \wse\ $\gamma$-ray emitting blazars 
(the \emph{Strip}) in the three \wse\ color-color planes~\cite{massaro12a}.
In this paper, we define the three-dimensional \emph{locus} by means of a 
Principal Component (PCs) analysis of the colors distribution of a
large sample of blazars composed by all the \bzcat\ sources with counterparts in the \wse\ 
All-Sky Catalog and associated to $\gamma$-ray source in the second \fer\ LAT catalog (2FGL) 
(the \wse\ \fer\ Blazars sample, WFB). Our new procedure yields a total completeness of 
$c_{\mathrm{tot}}\!\sim$81\% and total efficiency of $e_{\mathrm{tot}}\!\sim$97\%. We also 
obtain local estimates of the efficiency and completeness as functions 
of the \wse\ colors and galactic coordinates of the candidate blazars. The catalog 
of all \wse\ candidate blazars associated to the WFB sample is also presented, 
complemented by archival multi-frequency information 
for the alternative associations. Finally, we apply the new association 
procedure to all $\gamma$-ray blazars in the 2FGL  and provide a catalog 
containing all the $\gamma$-ray candidates blazars selected according to 
our procedure.

\end{abstract}


\section{Introduction}
\label{sec:intro}

Unveiling the nature of the Unidentified Gamma-ray Sources (UGS) 
is one of the main scientific objectives of the ongoing \fer\ $\gamma$-ray mission.
Recently, several attempts have been performed to associate or characterize the UGSs, either 
using X-ray observations \citep[e.g.,][]{mirabal09a,mirabal09b} or with statistical 
approaches~\citep[e.g.][]{mirabal10,ackermann12}. Nevertheless, according to~\cite{nolan12}, 
31\% of the $\gamma$-ray sources in the second \fer\ LAT catalog (2FGL) remain unidentified
and many of these unidentified sources could be blazars,
since blazars are known to dominate the $\gamma$-ray sky~\citep[e.g.][]{hartman99,abdo10a},
and among the 1297 associated sources within the 2FGL, 805 (62\%) are known blazars~\cite{nolan12}.
Therefore it is important to devise an efficient means of identifying candidate blazars 
among these sources.

Blazars are radio-loud Active Galactic Nuclei (AGNs) characterized by 
high and variable polarization, apparent 
superluminal motions, and high luminosities~\citep[e.g.,][]{urry95}. 
They exhibit a flat radio spectrum that steepens toward the infrared-optical bands. They also
show rapid variability from the radio to $\gamma$-rays and have peculiar infrared 
colors \citep{massaro11a}. Their characteristic spectral energy distributions (SEDs) 
have two main components: the low-energy component peaking in the spectral range 
from the IR to the X-ray band and the high-energy component peaking from MeV to TeV energies.

Blazars come in two main classes: the BL Lac objects, 
which have featureless optical spectra, and the more luminous 
Flat-Spectrum Radio Quasars which, typically, show prominent optical spectral emission
lines \citep{stickel91,stoke91}.
In the following discussion, we label the BL Lac objects as BZBs and the 
Flat-Spectrum Radio Quasars as BZQs, following the nomenclature 
of the Multi-wavelength Catalog of blazars \citep[\bzcat,][]{massaro09}.

Data from \wse\ have been recently used to select and classify AGNs~\citep{stern2012,yan2013} 
from a general point of view. The specific problem of the classification and identification of 
$\gamma$-ray sources has been tackled using machine learning techniques~\citep{hassan2013} 
and characterizing their optical variability~\citep{ruan2012}. Additional attempts to find 
counterparts of UGSs have been carried out using radio follow up 
observations~\citep[e.g.,][]{kovalev2009a,kovalev2009b}.

Using the preliminary data release of the Wide-field Infrared Survey Explorer (\wse) 
\citep[see][for more details]{wright10}\footnote{http://wise2.ipac.caltech.edu/docs/release/prelim/}, 
we showed that the $\gamma$-ray blazar population occupies a distinctive region 
of the \wse\ color parameter space (called the \wse\ Gamma-ray \emph{Strip}
\cite{massaro11a,dabrusco12}; hereinafter, Papers I and II respectively).

We then used the results of our analyses to develop a new association 
method to investigate the nature of the unidentified $\gamma$-ray sources in the 2FGL
\citep[][Paper III and IV]{massaro12a,massaro12b} as well as the unidentified 
{\it INTEGRAL} sources \citep[][Paper V]{massaro12c}.
Adopting our new association procedure, we have succeeded in finding low-energy 
counterpart candidates for 156 out of 313 ($\sim$50\%) of the unidentified 
$\gamma$-ray sources analyzed (Paper IV).

Taking advantage of the much larger data set now available thanks to the \wse\ 
All-Sky archive\footnote{http://wise2.ipac.caltech.edu/docs/release/allsky/}, 
released in March 2012, in this work we present a revisited definition of 
the region occupied by the $\gamma$-ray blazars. 
We will refer to the 3-dimensional region occupied by the $\gamma$-ray emitting blazars 
as the \emph{locus} while we will continue to indicate the 2-dimensional projection of the 
\emph{locus} in the $[3.4]\!-\![4.6]$ vs $[4.6]\!-\![12]$ $\mu$m color-color plane  as \wse\ Gamma-ray 
\emph{Strip}.
In this paper we determine a geometrical description of the WFB \emph{locus} in the space generated 
by the Principal Components of their distribution in the \wse\ color space and apply our association 
procedure the whole sample of blazars that belong to the 2FGL.

The paper is organized as follows: in Section~\ref{sec:sample} we describe the selection criteria 
adopted to create the samples needed for our investigation, together with the method 
for the positional associations between the \bzcat\ and the \wse\ archive. In 
Section~\ref{sec:locus} we describe the new geometric model of the WFB \emph{locus}
in the PCs space and define the quantitative parameter used to measure the compatibility
of a generic \wse\ source with the \emph{locus}.
Section~\ref{sec:association} is devoted to the association procedure 
based on the new parametrization of the \emph{locus}. In Section~\ref{sec:associationwfb}
and Section~\ref{sec:association2fb} we describe the results of the application of the new
association procedure to the WFB sample (with the estimates of the completeness and efficiency
of the process) and to the sample of 2FGL {\it Fermi} blazars, respectively. 
Section~\ref{sec:efficomp} discusses the evaluation of the efficiency and completeness
of the association procedure based n the results of the re-association of the WFB sample
presented in Section~\ref{sec:associationwfb}. In Section~\ref{sec:summary} we summarize 
the results of this paper and draw some conclusions. 

The \wse\ magnitudes in the [3.4], [4.6], [12], [22] $\mu$m nominal \wse\ filters 
are in the Vega system. We indicate the \wse\ colors [3.4]-[4.6], [4.6]-[12] and 
[12]-[22] as $c_{1}$, $c_{2}$ and $c_{3}$ and their corresponding errors as 
$\sigma_{c_{1}}$, $\sigma_{c_{2}}$ and $\sigma_{c_{3}}$, 
respectively. The values of the first two \wse\ magnitudes, namely $[3.4]$ 
and [4.6] and, in turn, of the two derived colors $c_{1}$ and $c_{2}$, have been 
corrected for galactic extinction according to the extinction law presented 
in~\cite{draine2003}. The corrections on the remaining two \wse\ magnitudes 
([12] and [22]) are negligible. All color values in the paper are corrected for 
galactic extinction.
All the acronyms used in the paper are listed in 
Table~\ref{tab:acronym}.

\begin{table}
	\caption{List of acronyms and specific terms used in this paper.}
	\begin{tabular}{lc}
	\tableline
	Name 									& 		Acronym 			\\
	\tableline\tableline
	Multiwavelength Catalog of blazars 				& 		\bzcat\ 			\\ 
	Second \fer\ Large Area Telescope catalog 		& 		2FGL 			\\
	Second \fer\ LAT Catalog of AGNs 				& 		2LAC 			\\
	\tableline
	BL Lac object 								& 		BZB 				\\
	Flat Spectrum Radio Quasar 					& 		BZQ 				\\
	Blazar of Uncertain type 						& 		BZU 				\\
	\wse\ {\it Fermi} Blazars sample 				& 		WFB 			\\
	2D region of the $c_{2}$ vs $c_{1}$ color-color	&						\\
	plane occupied by the WFB sample				&	\emph{strip}			\\ 
	3D region of the \wse\ color space 				&						\\
	occupied by the WBF sample 					& 	\emph{locus}			\\
	{\it Fermi} 2FGL Blazars sample				&		2FB				\\
	Search Region	for candidate blazars			&		SR				\\
	Background Region for candidate blazars		&		BR				\\	
	\tableline
	Principal Components 						&	 	PCs				\\
	\tableline
	\end{tabular}\\
	\label{tab:acronym}
\end{table}

\section{The sample selection}
\label{sec:sample}

\subsection{The ROMA-BZCAT}
\label{sec:bzcat}

The starting sample used in our analysis is the one presented in the \bzcat\ v4.1, released in August 2012, 
the most comprehensive blazars catalog existing in literature with 3149 \emph{bona fide} blazars and 
blazar candidates. The catalog includes 1220 BZBs, divided as 950 BL Lacs, 270 BL Lac candidates, 
1707 BZQs and 222 blazars of Uncertain type (BZUs) \citep{massaro11a}. All the details about 
the \bzcat\ catalog and the surveys used to build it 
can be found in \citep{massaro09,massaro10,massaro11b}\footnote{http://www.asdc.asi.it/bzcat/}.

\subsection{ \bzcat\ to \wse\ positional associations}
\label{sec:assoc}

We remark that the coordinates reported in the \bzcat\ are non-uniform. The astrometric accuracy 
can be as low as $\sim$ 5\arcsec, corresponding to the uncertainty of the NRAO Very Large Array 
Sky Survey (NVSS) for sources with radio fluxes close to the survey limit~\citep{condon98}.
For this reason, we have developed a method to perform the positional association between 
the \bzcat\ blazars and their \wse\ counterparts.
All \wse\ sources considered in our analysis are detected in at least one \wse\ band 
with a signal-to-noise ratio (SNR) higher than 5 in the \wse\ All-Sky .   

For each blazar, we searched for IR counterparts of the \bzcat\ blazars in the \wse\ all-sky archive
within circular regions of variable radius $R$ in the range between 0\arcsec and 6.5\arcsec. 
For each value of $R$, we estimated the number of total $N_t (R)$ and random matches $N_r (R)$, respectively,
together with the chance probability for the spurious associations, $p (R)$ calculated as follows.
The random matches $N_r (R)$ correspond to those found shifting the blazar 
coordinates by a fixed distance of 20\arcsec\ in a random direction of the sky.
This distance of 20\arcsec\ has been chosen because it is larger than the maximum positional uncertainty
on the coordinates reported in the \bzcat.
The chance probability for spurious associations $p (R)$ is calculated as the ratio between $N_r (R)$
and the total number of blazars listed in the \bzcat\ (i.e., 3149).

Then, we calculated the differences between the number of associations at given radius $R$ and those 
at $R-\Delta R$ for total matches, defined as:

\begin{eqnarray}
	\Delta N_t (R) &=& N_t(R)-N_t(R-\Delta R)~,
\end{eqnarray}

\noindent and the corresponding variation of the random associations, $\Delta N_r (R)$:

\begin{equation}
          \Delta N_r (R) = N_r(R)-N_r(R-\Delta R)~,
\end{equation} 

\noindent where $\Delta R$= 0.1\arcsec.

\begin{figure}[]
	\includegraphics[height=8cm,width=9cm,angle=0]{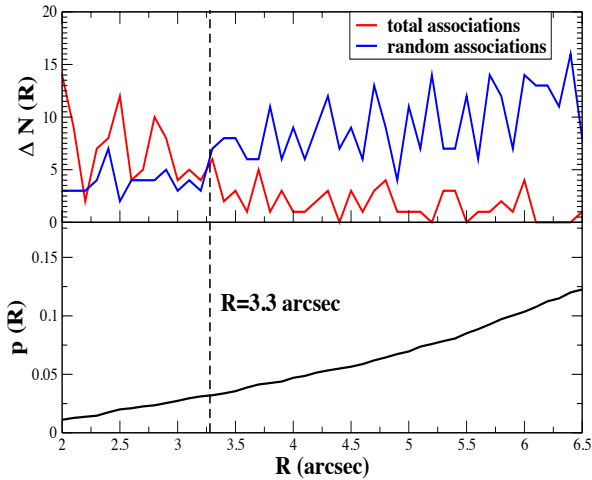}
         \caption{Upper panel) The number of total $N_t (R)$ (red line) and random 
          matches $N_r (R)$ (blue line) as function of the radius $R$ between 2\arcsec and 
          6.5\arcsec, respectively. 
          The radial threshold selected for our \bzcat\ - \wse\ crossmatches is indicated by
          the vertical dashed black line (see Section~\ref{sec:sample} for more details). 
          Lower panel) The chance probability for the spurious associations, $p(R)$ 
          as function of the radius $R$ between 2\arcsec ad 6.5\arcsec.}
          \label{fig:radii}
\end{figure}

Figure~\ref{fig:radii} shows the curves corresponding to $\Delta N_t (R)$, $\Delta N_r (R)$ and $p(R)$
for different values of the radius $R$ between 2\arcsec and 6.5\arcsec.
For all radii larger than 3.3\arcsec\ we found that the increase 
in number of IR sources positionally associated with blazars in the \bzcat\ becomes 
systematically lower than the increase in number of random associations; 
thus we chose this value as our radial threshold in searching for counterparts of 
\bzcat\ blazars in the \wse\ all-sky release. A similar positional association method 
has been adopted for the {\it INTEGRAL} sources \citep{stephen2006},
while the chance probability for the spurious associations, $p(R)$ has been estimated according to the 
procedure used for the {\it SWIFT} blazars detected in the hard X-rays by the BAT instrument on board
\citep{maselli10a,maselli10b}. We followed the same recipe also for searching the \wse\ blazars 
counterparts in Paper I.

\subsection{The \wse\ {\it Fermi} Blazars sample}
\label{sec:wbsssample}

We found 3032 out of 3149 (i.e., 96.3\% of the \bzcat) blazars with an IR counterpart 
within 3.3\arcsec\ in the \wse\ All-Sky data archive.  
In this sample, there are only 2 multiple matches out of 3032 spatial associations, 
for which we used the IR data of the closest \wse\ source in the following analysis. 
The probability of a chance associations for these 3032 is $\sim\!$ 3.3\% (see Figure~\ref{fig:radii}),
implying that $\sim\!$ 100 sources associated within the above radius could be spurious associations.

Of these 3032 blazars, 1172 are BZBs, including 919 BL Lacs and 253 BL Lac candidates,
1642 are BZQs and 218 are BZUs. It is also worth noticing that all the blazars 
associated between the \bzcat\ and the \wse\ all-sky data release 
are detected in the first two filters at 3.4 and 4.6 $\mu$m.
We also checked the properties of the blazars that have not been associated or do 
not have a counterpart in the \wse\ all-sky catalog, and we found that they do not 
appear to have peculiar properties in the radio, optical or X-ray energy range on 
the basis of the data reported in the \bzcat.

Among the 3032 selected blazars, only 673 have a counterpart in the $\gamma$-rays according to  
the 2FGL and to the CLEAN sample presented in the second \fer\ LAT Catalog of active 
galactic nuclei \citep[2LAC;][]{ackermann11}. 637/673 (i.e., 94.7\%) of these blazars (333 BZBs, 277 
BZQs, and 27 BZUs) are detected in all four \wse\ bands. As in Paper III the sample of $\gamma$-ray
emitting blazars in the \bzcat\ catalog was derived 
excluding the BZUs sources from our sample of $\gamma$-ray loud blazars. For this reason, the final
WFB sample includes only 610 \wse\ sources out of 673 \wse\ counterparts.

\section{The \emph{locus} parametrization}
\label{sec:locus}

In paper III we characterized the region occupied by the $\gamma$-ray blazars in the \wse\ 
IR color space by considering the three different two-dimensional projections in the 
color-color planes built with the \wse\ filters separately. Taking advantage of the \wse\ 
All-Sky archive recently released, in this paper we refine our previous definition of the \emph{locus} 
and improve significantly our association procedure by modeling the \emph{locus} 
occupied by the $\gamma$-ray blazars directly in the three dimensional parameter space generated
by the four \wse\ filters. This new parametrization and the improved association procedure
replace our previous analyses.

\subsection{The \emph{locus} in the Principal Components space}
\label{subsec:model}

\begin{figure}[] 
	\includegraphics[height=9cm,width=9cm,angle=0]{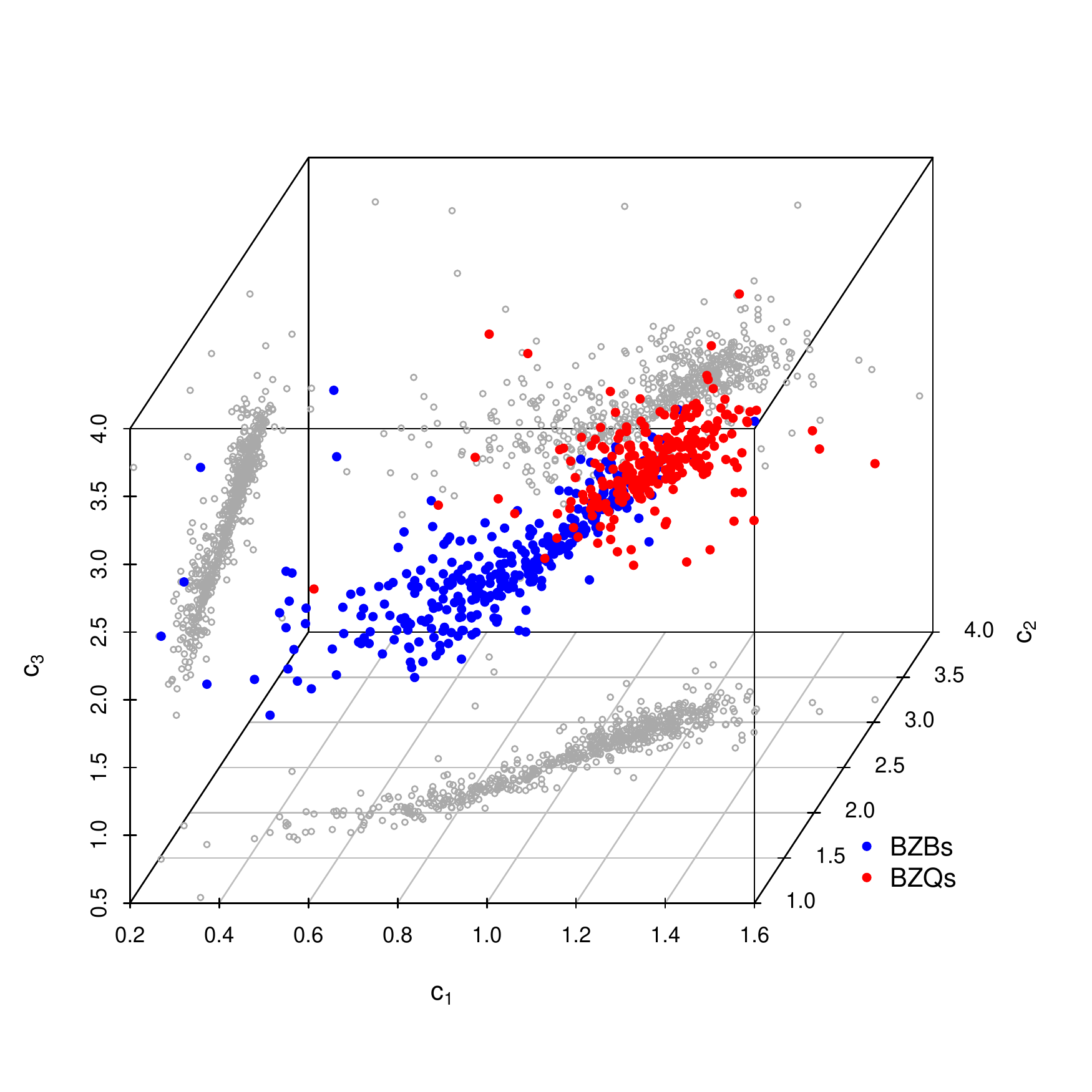}
	\caption{Scatterplot of the WFB sources in the three-dimensional \wse\ color space. The spectral 
	class of the WFB sources is color-coded, while the three distributions of gray points represent the 
	projections of the WFB sample in the three-dimensional color space onto the three two-dimensional 
	color-color planes generated by the \wse\ colors $c_{1}$, $c_{2}$ and $c_{3}$ respectively.}
          \label{fig:3d}
\end{figure}

The new parametrization of the WFB \emph{locus} is based on a new model of the \emph{locus} in the 
PCs space generated by the \wse\ colors of the WFB \wse\ counterparts, and 
on a more versatile definition of the statistical quantity used to evaluate the compatibility of a generic \wse\ 
source with the \emph{locus} model, the score. The distribution of the WFB 
sources in the three-dimensional \wse\ color space is axisymmetric along a slew line 
(see Figure~\ref{fig:3d}), so that a simple geometrical description of the \emph{locus} can be 
determined in the PCs space. 

Principal Component Analysis (PCA) uses an 
orthogonal transformation to convert a set of observations of possibly correlated 
variables into a set of values of linearly uncorrelated variables, the PCs. This 
transformation $T\!:\!(c_{1},c_{2},c_{3})\rightarrow(\mathrm{PC}_{1},\mathrm{PC}_{2},
\mathrm{PC}_{3})$ is defined so that the first PC (PC$_{1}$) accounts for as much of the variance in 
the data as possible, and each following component (PC$_{2}$, PC$_{3}$, etc. up to 
the dimensionality of the initial space) has the 
highest variance possible under the constraint that it is orthogonal to the preceding components. 
In our case, the WFB
sources in the three-dimensional PCs space based on their color distributions lie almost 
perfectly along the PC$_{1}$ axis and are distributed symmetrically in the PC$_{2}$ vs PC$_{3}$ 
plane around the PC$_{1}$ line. Based on the shape of the \emph{locus} in the PCs space, 
we choose to define its geometrical model using a cylindrical parametrization, 
with axis aligned along the PC$_{1}$ axis. 
The \emph{locus}, as a whole, is modeled by three distinct cylinders: the first two of these cylindrical 
regions are dominated by BZB and BZQ sources respectively, while the third cylinder is defined as the 
region where the WFB population is \emph{mixed} in terms of spectral classes (the Mixed region, 
hereinafter). 

\begin{figure}[] 
	\includegraphics[height=9cm,width=9cm,angle=0]{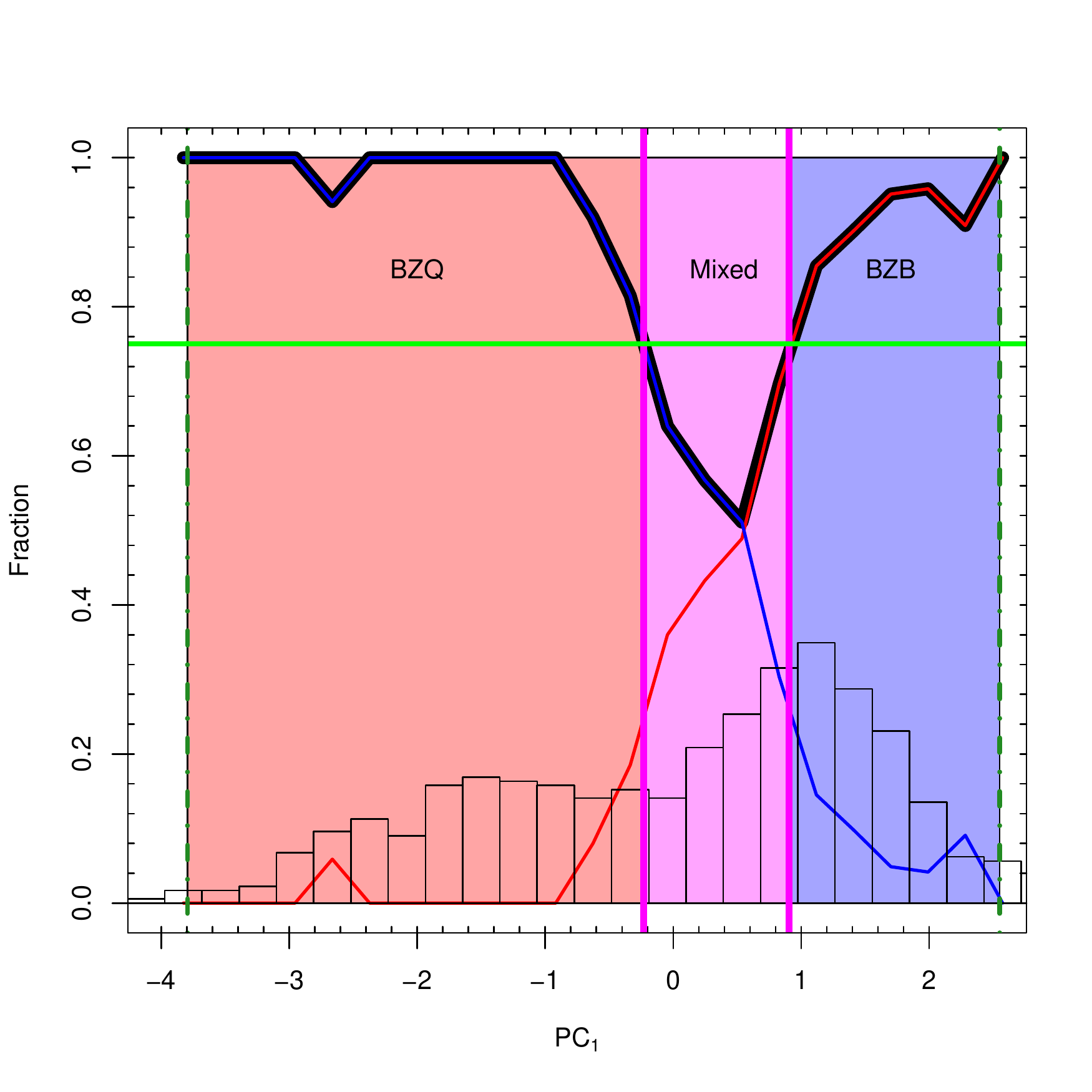}
          \caption{Boundaries of the three sections of the WFB \emph{locus} in the PCs space
          along the PC$_{1}$ axis. The solid black line represent the ``purity'' of the WFB population, 
          i.e. the fraction of the dominant spectral class relative to the other spectral class. The solid 
          red and blue lines represent the fraction of BZQs and BZBs sources, while the histogram 
          in the background represents the normalized density of the distribution of the whole WFB sample
          along the PC$_{1}$ axis. The horizontal 
          green line shows the threshold used to determine the boundaries of the mixed region.}
          \label{fig:cylinders}
\end{figure}

The upper and lower boundaries of the model along the 
PC$_{1}$ axis have been determined requiring that 90\% of the total number of WFB sources is contained 
within the boundaries of the cylinder, with 5\% of the sources outside of the boundaries of the model 
on each side of the model along the PC$_{1}$ axis. 
The boundaries of the Mixed section along 
the PC$_{1}$ axis have been defined by requiring 
that, in this region, the fraction of either spectral class is smaller than 80\% of the total number of WFB 
sources. The three boundaries along the PC$_{1}$ axis defining the three sections 
of the WFB \emph{locus} model are shown in Figure.~\ref{fig:cylinders}.

\begin{figure}[] 
	\includegraphics[height=9cm,width=9cm,angle=0]{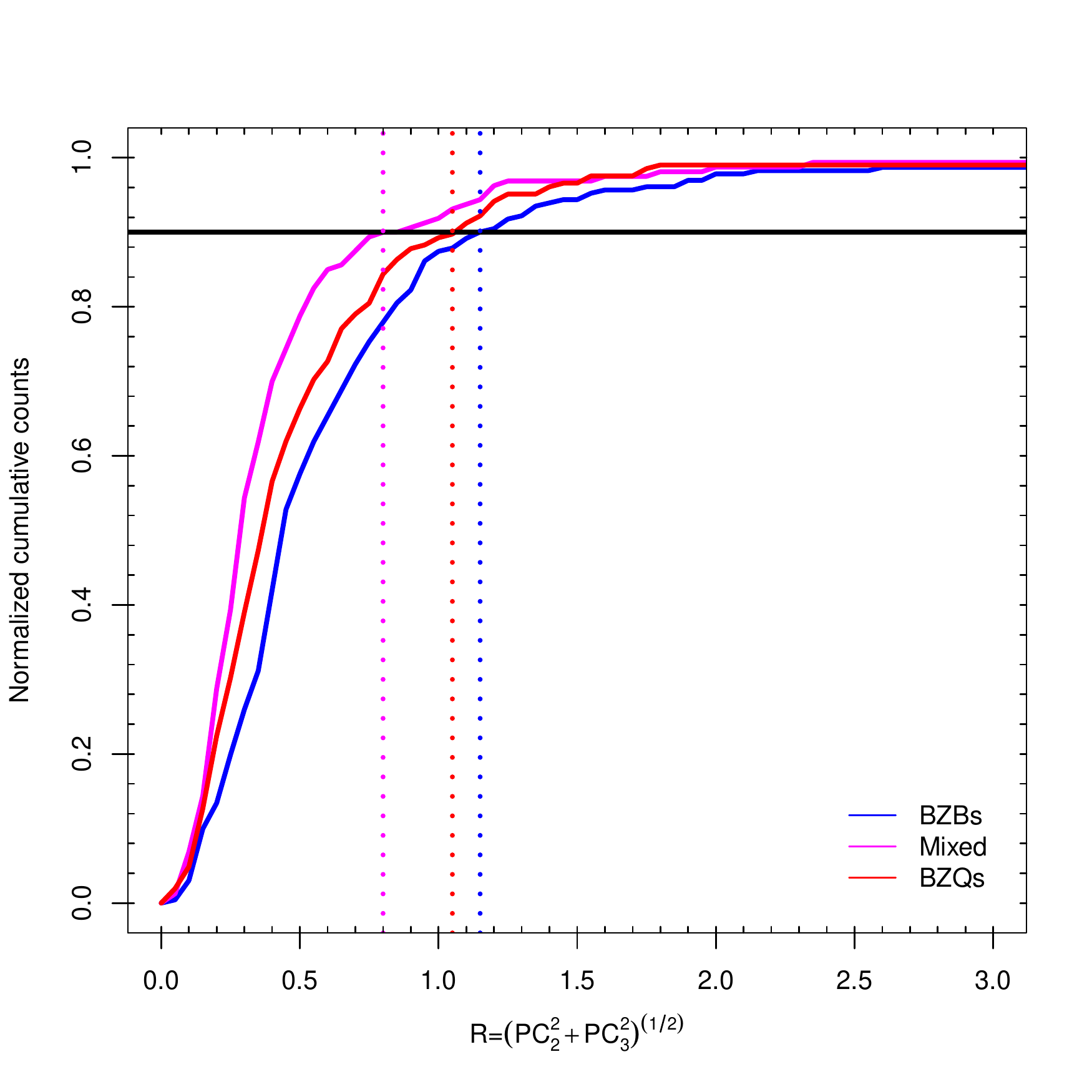}
          \caption{Cumulative radial profiles of the WFB sources for the three different sections of the 
          model of the WFB \emph{locus} in the PCs space. The red, magenta and blue solid lines represent
          the cumulative profiles of the WFB sources belonging to the BZQs, mixed and BZBs sections of 
          the model, respectively. The horizontal black line shows the 90\% threshold used to determined 
          the radii of the three cylinders (dotted vertical lines).}
          \label{fig:cylindersradii}
\end{figure}

\noindent The variances of the distribution of the WFB distribution in the
PC space along the second and third PCs are $\sigma^2_{\mathrm{PC_{2}}}\!=\!0.61$ and 
$\sigma^2_{\mathrm{PC_{3}}}\!=\!0.58$ respectively. Based on this fact, we have 
modeled the bases of the cylinders as circles centered on the axis of the first principal component
PC$_{1}$ (the variance of the WFB distribution along PC$_{1}$ is 
$\sigma^1_{\mathrm{PC_{3}}}\!=\!1.53$). 
The radii of the circular bases of each of the three cylinders representing the three
different sections of the WFB \emph{locus} in the PCs space have been determined independently as 
the radii containing the 90\% of the WFB sources in each section. The radii of each of the three 
cylinders are defined 
in the plane generated by the PC$_{2}$ and PC$_{3}$ axes and evaluate d as  
$R=\sqrt{\mathrm{PC}_{2}^2\!+\!\mathrm{PC}_{3}^2}$. The cumulative radial profiles of the three
sections and the corresponding radii determined as discussed above are shown in 
Figure~\ref{fig:cylindersradii}. The numerical values of the boundaries along the PC$_{1}$ axis
and the radii of the three cylinders of the new model of the WFB \emph{locus} are reported in 
Table~\ref{tab:model}. 

\begin{table}
	\begin{center}
	\caption{Parameters of the three cylinders representing the model of the WFB \emph{locus}
	in the PCs space.}
	\begin{tabular}{cccc}
	\tableline\tableline
				& BZB 	& Mixed	& BZQ	\\
	\tableline
	PC$_{1}$ low.	& -3.79	& -0.24	& 0.92	\\
	PC$_{1}$ up.	& -0.24	& 0.92	& 2.55	\\
	radius		& 1.15	& 0.8		& 1.05	\\	
	\tableline
	\label{tab:model}
	\end{tabular}
	\end{center}
\end{table}

\subsection{The score}
\label{subsec:score}

\begin{figure}[] 
	\includegraphics[height=9cm,width=9cm,angle=0]{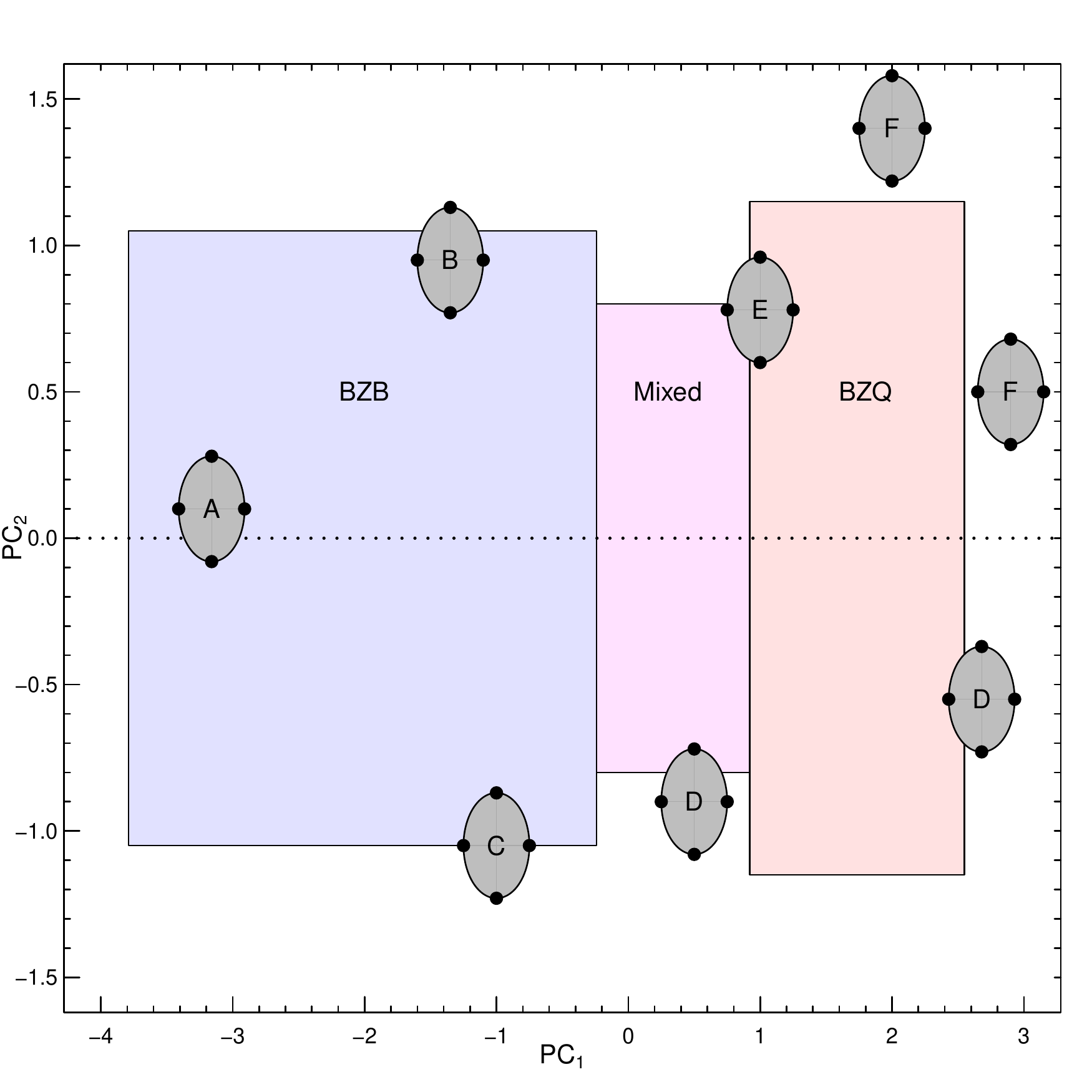}
           \caption{Schematic representation of the possible positions of the uncertainty ellipsoid
           of a generic \wse\ source in the PCs space relative to the cylindrical models of the WFB
           \emph{locus}. In this plot the projection of the three-dimensional model and the uncertainty 
           ellipsoids on a plane containing the axis of the cylinders are shown. The letters are references
           to the description of the different cases in Sec.~\ref{subsec:score}.}
           \label{fig:scheme}
\end{figure}

The distance of a generic \wse\ source to the model of the WFB \emph{locus} in 
the PCs space can be evaluated quantitatively using a numeric quantity that we call the score. 
The generic \wse\ source with colors $(\tilde{c}_{1}, \tilde{c}_{2}, \tilde{c}_{3})$ can be projected onto
the PCs space by applying the orthogonal transformation determined by the PCA performed on 
the WFB sample for the modelization of the WFB \emph{locus} in the PCs space (Section~\ref{subsec:model}). 
Thus, the position of the generic \wse\ source in the PCs space is determined by the 
PCs values $(\tilde{\mathrm{PC}}_{1},\!\tilde{\mathrm{PC}}_{2},
\!\tilde{\mathrm{PC}}_{3})\!=\!T(\tilde{c}_{1},
\!\tilde{c}_{2},\!\tilde{c}_{3})$. To take into account the uncertainties on the values of the \wse\ colors,
the standard deviations on each color are also projected onto the PCs space and are used to 
define the error bars on the position of the source in the PCs space: $(\pm\sigma_{\tilde{\mathrm{PC}}_{1}},
\!\pm\sigma_{\tilde{\mathrm{PC}}_{2}},\!\pm\sigma_{\tilde{\mathrm{PC}}_{3}})\!=\!T(\pm\sigma_{\tilde{c}_{1}},
\!\pm\sigma_{\tilde{c}_{2}},\!\pm\sigma_{\tilde{c}_{3}})$. We simply assume that the 
generic \wse\ source is represented in the PCs space by the ellipsoid generated by the 
segments with extremes $\mathrm{PC}_{i} \pm \sigma_{\mathrm{PC}_{i}}$, hereinafter the 
uncertainty ellipsoid. Each of the six points at the extremes of the axes of the uncertainty 
ellipsoid in the PCs space will be generically called extremal point. The possible
positions of the uncertainty ellipsoid associated with a generic \wse\ source relative to each 
of the three cylinders of the \emph{locus} model
(schematically shown in Figure~\ref{fig:scheme} for one two-dimensional section of the PCs space) 
fall in one of the following cases:

\begin{itemize}
	\item Six extremal points within a cylinder (point A in Figure~\ref{fig:scheme});
	\item Five extremal points within a cylinder (point B in Figure~\ref{fig:scheme}); 
	\item Three extremal points within a cylinder (point C in Figure~\ref{fig:scheme});
	\item One extremal point within a cylinder (points D in Figure~\ref{fig:scheme});
	\item No extremal points within any cylinder (points F in Figure~\ref{fig:scheme});
\end{itemize}

\noindent Other combinations are not possible because the axes of the uncertainty 
ellipsoids are either parallel or orthogonal to the PC$_{1}$ axis of the PCs space. Points with
any number of extremal points within two cylinders (like point E in Figure~\ref{fig:scheme}) are assigned
a distinct score value for either cylinder according to the number of extremal points contained
in each one.
 			
\noindent The score $s$ for a generic \wse\ source with $n$ extremal points contained in one of the three
sections of the WFB \emph{locus} model is then defined as:

\begin{equation}
	\nonumber
	\mathrm{s} = \frac{1}{6^{\phi}}\cdot\!n^{\phi}
	\label{eq:score}
\end{equation}

\noindent where $\phi$ is the \emph{index} of the score assignment law. 
This is a simple generalization of the most natural choice that would assign to
each extremal point within the \emph{locus} model 1/6, defining the total score of a
source as linearly proportional to the number of extremal points within the model cylinders. 
This behavior is obtained in the general equation when $\phi\!=\!1$. 
Changing the value of $\phi$ is useful to tweak the performances of the association 
procedure in terms of the purity and completeness of the final sample of candidate blazars.
While in this paper $\phi\!=\!1$ will be used for all experiments, the discussion of the 
influence of the value of the parameter $\phi$ on the results of the 
application of the association procedure and the justification of  
the choice of the value of $\phi$ used in the experiments in this paper 
can be found in Appendix A.

So far, the score assigned to a generic \wse\ source can take one of six different 
values determined by the score assignment law in Equation~\ref{eq:score}. To 
penalize the \wse\ sources with large uncertainties on the observed colors (and, 
in turn, large volume of the uncertainty ellipsoid in the PCs space) relatively to 
other \wse\ sources with the same number of extremal points contained in the 
\emph{locus} model but smaller errors, we multiply the score obtained using 
Equation~\ref{eq:score} by the
ratio of the absolute values of the logarithms of the volume of the uncertainty 
ellipsoid of the source considered and of the volume of the largest uncertainty ellipsoid
for WFB sources. Thus, for each of the three regions of the \emph{locus} model, the 
weighted score is defined as:

\begin{equation}
	\nonumber
	\mathrm{s}_{w} = \mathrm{s}\!\cdot\frac{\|\log{V}\|}{\|\log{(\mathrm{max(V_{\mathrm{WFB}})})}\|}
	\label{eq:weiscore}
\end{equation}
 
\noindent where $V_{\mathrm{WFB}}$ are the volumes of the uncertainty ellipsoids of the 
WFB sources in the PCs space calculated as 
$V_{\mathrm{WFB}}\!=\!\frac{4}{3}\pi\sigma_{\mathrm{PC}_{1}}\sigma_{\mathrm{PC}_{1}}\sigma_{\mathrm{PC}_{3}}$, 
and $V$ is the volume of the uncertainty ellipsoid in the 
PCs space of the generic \wse\ source considered. The logarithms of the 
volumes of the uncertainty ellipsoids are used to take into account the 
large number of order of magnitude potentially spanned by the differences 
between the volumes (always smaller than one in the PCs space though).
The above definition of the weighted score also has the effect of mapping 
the discrete distribution of scores calculated according to assignment law 
Equation~\ref{eq:score} into a continuous distribution that allows a finer 
classification of the candidate blazars (as described in Section~\ref{subsec:candidates}).
In the remainder of the paper the term score will always be used to indicate
the weighted score definition given in Equation~\ref{eq:weiscore}.

\section{The association procedure}
\label{sec:association}

\subsection{Selection of the candidate blazars}
\label{subsec:candidates}

\begin{figure}[] 
          \includegraphics[height=9cm,width=8.5cm,angle=0]{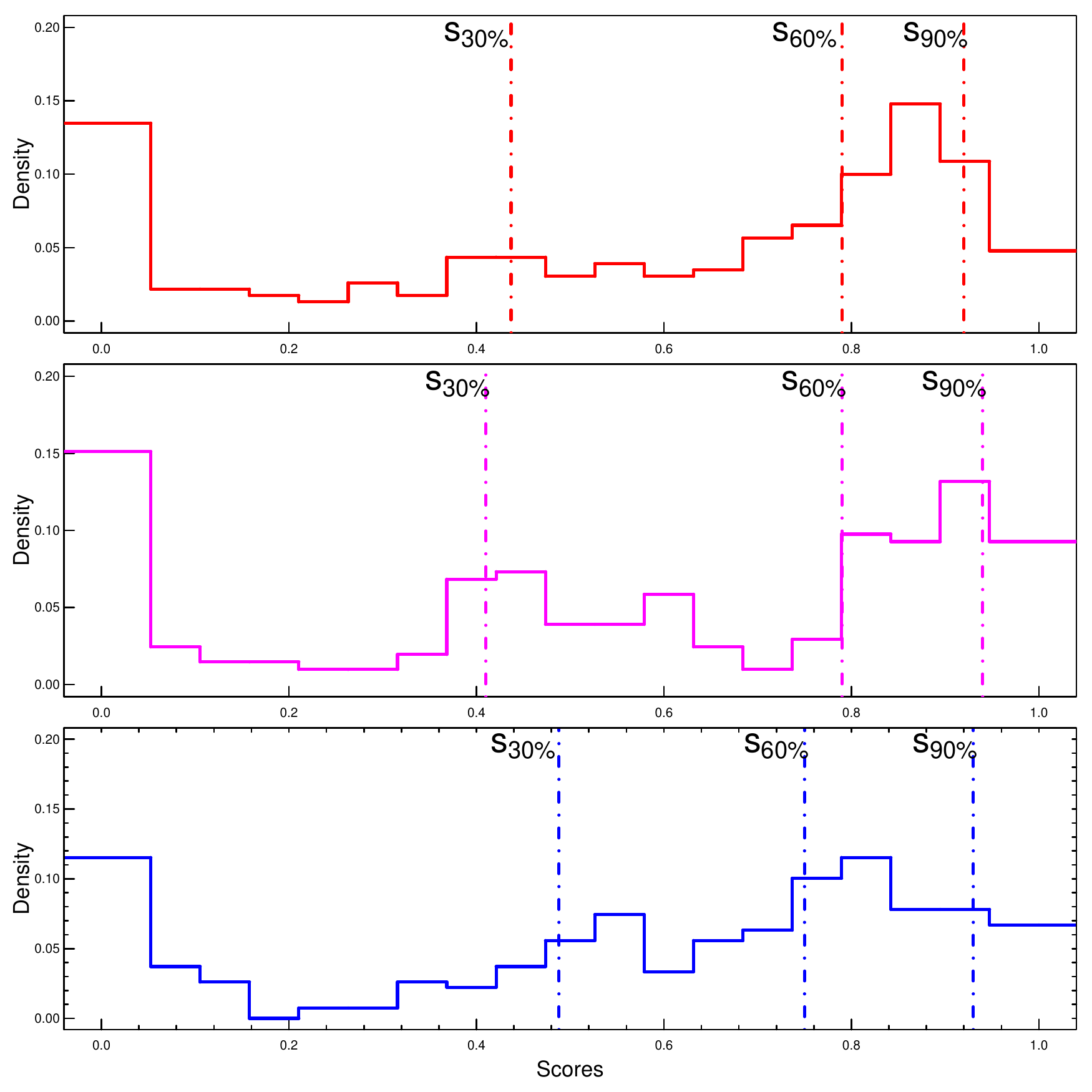}
          \caption{Histograms of distributions of score values calculated for the sources in the WFB sample 
          for the three regions of the \emph{locus} dominated respectively by the BZQs, the BZBs and
          in the mixed region (upper, mid and and lower panels respectively). The three 
          vertical lines in each panel represent $s_{30\%}$, $s_{60\%}$ and $s_{90\%}$, i.e. the values of 
          the score associated with the 30\%-th, 60\%-th and 90\%-th
          percentiles for BZBs, BZQs and mixed sources respectively. These score thresholds have
          been used to define the classes of candidate blazars (see Section~\ref{sec:association} 
          for details).}
          \label{fig:scoreswgs}
\end{figure}

The procedure for the evaluation of the scores based on the new parametrization of the WFB \emph{locus} 
discussed in the previous section is used to associate
high-energy sources to \wse\ candidate blazars. The \wse\ colors and their uncertainties for all 
the sources found in the \wse\ All-Sky photometry catalog within the region of positional 
uncertainty (hereinafter the Search Region - SR) of a given high-energy source and 
detected in all four \wse\ filters are retrieved, and the scores of these \wse\ sources are calculated
as described in Section~\ref{subsec:score}.
Then, these sources are split among different classes 
according to the values of the their scores
$s_{b}$, $s_{m}$ and $s_{q}$ for the BZB, Mixed and BZQ regions of the WFB \emph{locus} model
in the PCs space respectively. For each \emph{locus} region, every source 
is assigned to class A, class B, class C or is marked as an outlier based on its score values and relative 
to the threshold scores values defined as the 30\%, 60\% and 90\% percentiles of the distributions 
of scores in the three regions of the \emph{locus} for the WFB sources (see Figure~\ref{fig:scoreswgs}). 
The classes are sorted according to decreasing probability of the \wse\ source to be compatible with 
the model of the WFB \emph{locus}: class A sources are considered the most probable candidate 
blazars for the high-energy source in the SR, while class B and class C sources are less compatible
with the WFB \emph{locus} but are still deemed as candidate blazars. In more details, 
class A candidate blazars have score $s\leq s_{90\%}$, class B candidate blazars have score 
$s_{60\%}\!\leq\!s\leq\!s_{90\%}$ and class C candidate blazars have score 
$s_{30\%} \leq s\leq s_{60\%}$ for each region. The other sources considered 
outliers are discarded. The values of the score thresholds derived from 
the score distributions of WFB sources for the 
three regions of the \emph{locus} model are reported in Table~\ref{tab:thresholds} and shown 
in Figure~\ref{fig:scoreswgs} overplotted to the histograms of the score distributions of the WFB sources
assigned to each of the three \emph{locus} regions.

\begin{table}
	\begin{center}
	\caption{Values of the score thresholds $s_{30\%}$, $s_{60\%}$ and $s_{90\%}$, used for 
	the association experiments described in this paper. These values are determined as the 
	30\%-th, 60\%-th and 90\%-th percentiles
	of the scores of the WFB sample divided by BZB, Mixed and BZB mixed regions.}
	\begin{tabular}{cccc}
	\tableline\tableline
					& BZB 	& Mixed	& BZQ  	\\
	\tableline
	$s_{30\%}$		& 0.48	& 0.44	& 0.41	\\
	$s_{60\%}$		& 0.75	& 0.79	& 0.79	\\
	$s_{90\%}$		& 0.93	& 0.92	& 0.94	\\	
	\tableline
	\label{tab:thresholds}
	\end{tabular}
	\end{center}
\end{table}

\noindent The choice of the percentiles used to define the classes of candidate blazars is 
arbitrary and can be changed to allow for more conservative (higher purity of the sample
of candidates) or more complete (lower purity of the sample of candidates) selections of 
candidate blazars in the SRs associated with unidentified high-energy sources. 

\subsection{Background and spurious associations}
\label{sec:back}

\begin{figure}[] 
	\includegraphics[height=9cm,width=9cm,angle=0]{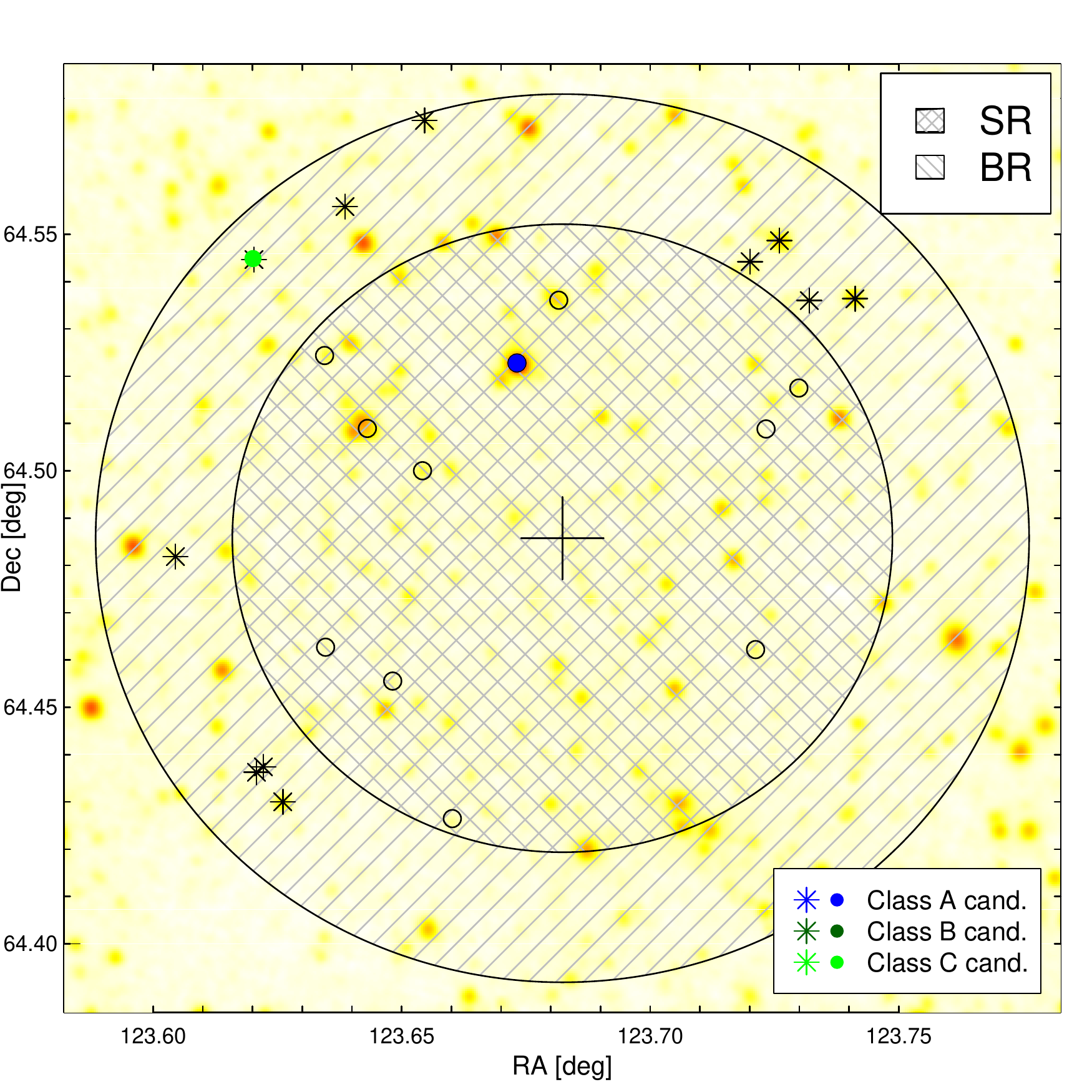}
           \caption{Results of the association procedure for a generic unassociated high-energy source
           superimposed on the image of the \wse\ sky around the position of the unassociated $\gamma$-ray
           source as seen in the $[3.4]\mu$m band. 
           The inner circle represents the Search Region (SR) of the high-energy source while the outer 
           circle delimits the annulus used as Background Region (BR). The open circles in the SR represent 
           the sources of the \wse\ All-Sky catalog 
           detected in all four \wse\ filters for which the scores have been evaluated (the sources not marked 
           by symbols in the image are not detected in at least one of the four \wse\ filters and have not been
           considered for the score evaluation). The solid circle
           represents the candidate blazar found within the SR and its color indicates that it is class A candidate
           blazar. The gray stars in the 
           background region are the \wse\ sources used to assess the possibility of spurious associations.
           Only one source in the BR has been classified as class C candidate blazar in this example.}
          \label{fig:association}
\end{figure}

In our association procedure, the presence of \wse\ background sources with score values
that would qualify them as candidate blazars but that are not located within the SR of the  
unidentified high-energy source is taken into account by assessing
the number and type of spurious associations from sources within a local background 
region for each unassociated source.
For a generic SR of radius $r_{\mathrm{SR}}$, we define the background region (BR) as an 
annulus of outer radius $r_{\mathrm{BR}}\!=\!\sqrt{2}\!\cdot\!r_{\mathrm{SR}}$ and inner
radius equal to the SR radius and centered on the center of the SR. The SR and BR have same
area by definition. Within a given SR, all \wse\ sources detected in all four \wse\ filters are 
assigned a score value for each 
region of the \emph{locus} model, and successively ranked in classes using the same thresholds
used to classify the sources within the SR. An example of a generic SR and associated 
background region is shown in Figure~\ref{fig:association}, where the candidate blazar and 
the spurious BR candidate blazar are colored according to their class membership as defined in 
Section~\ref{sec:association}.

\noindent For every unassociated high-energy source, our method produces all
candidate blazars (sources classified as class A, class B or class C candidate) in the SR. 
All candidate blazars located in the BR of the high-energy sources are also provided 
and can be used to evaluate the chance of spurious associations as a function of the class 
of the candidate blazars (see, for example, Section~\ref{sec:associationwfb} and 
Section~\ref{sec:association2fb} for WFB and 2FB association respectively, and the last 
columns of Table~\ref{tab:wfbcatalogsamplereassociated} and Table~\ref{tab:catalogsample2fb}). 

\section{Re-association of the \wse\ $gamma$-ray emitting blazars}
\label{sec:associationwfb}

The association procedure based on the new model of the \emph{locus} of the WFB in the 
3-dimensional \wse\ color space (described in Sec.~\ref{subsec:model}) has been applied to 
the $\gamma$-ray sources of the WFB sample. The candidate blazars
have been classified according to the score thresholds described in Sec.~\ref{sec:association} and shown
in Table~\ref{tab:thresholds}. The SR associated with each $\gamma$-ray source has been defined as a
circular region of radius $\theta_{95}$ corresponding to the semi-major axis of the uncertainty
positional region at 95\% level of confidence \citep[see][for additional details]{nolan12}.
The method finds 542 candidate blazars for 486 out of 610 WFB 
$\gamma$-ray sources, corresponding to $\sim80\%$ of the WFB sample. 
The 542 candidate blazars are divided into 186 BZB candidate blazars, 165 candidate blazars 
compatible with the Mixed region and the remaining 191 BZQ candidate
blazars. The 542 candidate blazars are distributed in 
75 class A candidates, 211 class B candidates and 256 class C candidate blazars. The average 
number of candidate blazars for each associated WFB $\gamma$-ray source is $\sim$1.1, with 197 
WFB sources associated with only one \wse\ candidate. 
Out of the total of 486 $\gamma$-ray sources in the WFB that have been associated with at least
one \wse\ candidate blazar, 468 sources have been associated (among other possible candidate blazars
selected) to the same \wse\ source which has been selected as the counterpart of the $\gamma$-ray 
source according to the analysis discussed in Section~\ref{sec:wbsssample}, corresponding to 
$\sim$77\% of the WFB sample. More details on this topic are given in Section~\ref{sec:efficomp} 
where the evaluation of the efficiency and completeness of the association procedure is discussed.
A summary of the composition of the sample of candidate blazars selected by the new association 
method applied to the WFB sample can be found in Table~\ref{tab:composition}.

\begin{table}
	\begin{center}
	\caption{Composition of the samples of candidate blazars selected by applying our 
	association method to the $\gamma$-ray sources in the WFB and 2FB samples in terms of 
	candidate types and class. See Section~\ref{sec:associationwfb} and 
	Section~\ref{sec:association2fb} for more details on the re-association of the WFB sample
	and the association of the 2FB sample respectively.}
	\begin{tabular}{cccc}
	\tableline\tableline
				& 	 		& WFB				& 		 	\\
				& Class A 		& Class B				& Class C 	\\
	\tableline
	BZB			& 28			& 81					& 77			\\
	Mixed		& 23			& 59					& 83			\\
	BZQ			& 24			& 71					& 96			\\
	\tableline
				& 	 		& 2FB				& 		 	\\
				& Class A 		& Class B				& Class C 	\\	
	\tableline
	BZB			& 0			& 13					& 5			\\
	Mixed		& 4			& 13					& 9			\\
	BZQ			& 4			& 8					& 20			\\	
	\tableline
	\label{tab:composition}
	\end{tabular}
	\end{center}
\end{table}

\noindent Table~\ref{tab:wfbcatalogsamplereassociated} shows the basic information for the 
first ten candidate blazars of the WFB sample that have been re-associated with the same \wse\ 
counterparts used in the WFB catalog itself. 
The complete list of candidate blazars associated with the WFB sources can be found in the 
Table~\ref{tab:wfbcatalogreassociated} in Appendix B.

\begin{table*}
	\begin{center}
	\begin{threeparttable}	
	\caption{First ten \wse\ candidate blazars associated to WFB sources and corresponding to the 
	association in the 2FGL catalog (the complete list of WFB associations can be found in Table~\ref{tab:wfbcatalogreassociated}
	in Appendix B).}
	\label{tab:wfbcatalogsamplereassociated}
	\begin{tabular}{lllcccccl}
	\tableline\tableline
  	2FGL\tnote{a}  		&  \wse\tnote{b}  	& \bzcat\tnote{c} 	& [3.4]-[4.6]\tnote{d} 	& [4.6]-[12]\tnote{e} 	& [12]-[22]\tnote{f} 		& z\tnote{g} 	& type\tnote{h} 		& class\tnote{i}  \\
  	name  			&  name   			&  name   			&     mag    	 	&    mag     		&   mag     				&   			&      				&    			\\
	\tableline
 	2FGLJ0000.9-0748 		& J000118.01-074626.7 		& BZBJ0001-0746 	& 0.93(0.03)	& 2.53(0.04) 		& 2.12(0.1)	& ?       	& BZB 	& B 	\\
  	2FGLJ0004.7-4736 		& J000435.64-473619.5 		& BZQJ0004-4736 	& 1.09(0.03)	& 2.92(0.03) 		& 2.3 (0.06) 	& 0.88   	& UND 	& C 	\\
  	2FGLJ0006.1+3821 		& J000557.17+382015.2 		& BZQJ0005+3820 	& 1.1 (0.03)	& 3.11(0.03) 		& 2.69(0.03) 	& 0.229   	& BZQ 	& A 	\\
  	2FGLJ0007.8+4713 		& J000759.97+471207.7 		& BZBJ0007+4712 	& 0.92(0.04)	& 2.28(0.06) 		& 2.13(0.18)	& 0.28    	& BZB 	& C 	\\
  	2FGLJ0012.9-3954 		& J001259.88-395425.8 		& BZBJ0012-3954 	& 1.01(0.04)	& 2.84(0.04) 		& 2.28(0.11) 	& ?       	& UND 	& C 	\\
 	2FGLJ0013.8+1907 		& J001356.36+191042.0 		& BZBJ0013+1910 	& 0.97(0.04)	& 2.58(0.06) 		& 2.09(0.22)	& ?       	& BZB 	& C 	\\
  	2FGLJ0017.6-0510 		& J001735.81-051241.6 		& BZQJ0017-0512 	& 1.06(0.03)	& 2.73(0.04) 		& 2.45(0.08) 	& 0.227   	& UND 	& B 	\\
  	2FGLJ0021.6-2551 		& J002132.54-255049.2 		& BZBJ0021-2550 	& 0.85(0.04)	& 2.3 (0.06) 		& 2.23(0.1) 	& ?       	& BZB 	& C 	\\
  	2FGLJ0022.5+0607 		& J002232.44+060804.4 		& BZBJ0022+0608 	& 1.04(0.03)	& 2.8 (0.04) 		& 2.28(0.09) 	& ?       	& UND 	& B 	\\
	2FGLJ0023.2+4454 		& J002335.44+445635.8 		& BZQJ0023+4456 	& 1.22(0.04)	& 2.92(0.06) 		& 2.41(0.14)	& 1.062   	& BZQ 	& C	\\
	\tableline
	\end{tabular}	
	\begin{tablenotes}[para]
 Ê Ê Ê Ê Ê	\item {Notes:}\\
 Ê Ê Ê Ê Ê	\item[a] 2FLG name of the $\gamma$-ray source associated\\
		\item[b] \wse\ name of the candidate blazars\\
		\item[c] \bzcat\ name of the source\\
		\item[d] $c_{1}$ \wse\ color of the candidate\\
		\item[e] $c_{2}$ \wse\ color of the candidate\\
		\item[f] $c_{3}$ \wse\ color of the candidate\\
		\item[g] redshift of the \bzcat\ source\\		
		\item[h] classification of the candidate blazar according to our method\\
		\item[i] class of the candidate blazar according to our method\\ 
	\end{tablenotes}
	\end{threeparttable}	
	\end{center}
\end{table*}

\noindent Table~\ref{tab:wgscatalogsamplealternative} shows some basic information 
for the first ten candidate blazars associated with the WFB sample but 
representing alternative associations of the $\gamma$-ray sources in the WFB sample (i.e., 
different \wse\ sources from the counterparts identified for the WFB sample with the 
positional method described in Section~\ref{sec:assoc}). For each of 
these candidate blazars, an extensive archival research has been performed in order to 
gather all critical information useful to characterize the nature of the source. 
This information is reported in the tenth column of Table~\ref{tab:wgscatalogsamplealternative}, 
summarizing the available classifications and detections in different observations for each
alternative association. 
The last column of this Table also reports the total number of BR
candidate blazars, determined as described in Section~\ref{sec:back}, with class higher than or 
equal to the class of the candidate blazars associated with the $\gamma$-ray source 
and located within its SR. No class A candidate blazars have been found within the 
BRs of the 75 WFB sources associated to at least one class A candidate blazar 
(0\%), while 32 out of the 201 WFB sources associated to a class B candidate
have at least one class B or better candidate blazar in the BR ($\sim\!16\%$). Finally, 
167 out of the remaining 192 WFB sources associated to a class C candidate blazar 
have at least one candidate blazar of equal or better class in their BRs ($\sim\!87\%$), 
confirming that class C candidate blazars are the most sensitive to possible contamination 
from background \wse\ sources compatible with the \emph{locus} model.  
The complete list of alternative associations of the WFB sources 
can be found in Table~\ref{tab:wfbcatalogreassociated} in Appendix B.

\begin{table*}
	\tiny
	\begin{center}
	\begin{threeparttable}	
	\caption{First ten candidate blazars associated to WFB sources that do not correspond to the \wse\ counterparts 
	of the associations in the 2FLG catalog (the complete list of WFB associations can be found in 
	Table~\ref{tab:wfbcatalogreassociated} in Appendix B.}
	\begin{tabular}{lllccccclcc}
	\tableline\tableline
  	2FGL\tnote{a}  	&  \wse\tnote{b}  	&  other\tnote{c} 	& [3.4]-[4.6]\tnote{d} 	& [4.6]-[12]\tnote{e} 	& [12]-[22]\tnote{f} 	& type\tnote{g} 	& class\tnote{h}& notes\tnote{i} & z\tnote{l} 	& $N_{\mathrm{BR}}$\tnote{m}			\\
  	name  		&  name   			&  name  			&     mag     		&    mag     		&   mag     			&      			&       		&       		&   			& 								\\
	\tableline
  	2FGLJ0000.9-0748 		& J000115.93-074233.1 		& APMUKS(BJ) B235842.09-075916.3 	& 1.03(0.05) & 3.21(0.08) & 2.45(0.20) & BZQ 		& C & -    		& 		& 1	\\
  	2FGLJ0007.8+4713 		& J000745.11+471130.7 		& NVSSJ000745+471131             		& 1.15(0.06) & 2.92(0.12) & 2.68(0.27) & BZQ 		& C & N,X 	& 		& 0	\\
  	2FGLJ0102.3+4216 		& J010142.98+421828.3 		& GALEXJ010142.98+421828.4       		& 1.00(0.05) & 2.76(0.09) & 2.40(0.28) & UND 		& C & -    		& 		& 1	\\
  	2FGLJ0116.0-1134 		& J011559.75-113012.3 		& GALEX2673671438794228469       	& 1.13(0.03) & 2.46(0.03) & 2.36(0.05) & UND 		& C & M   		&		& 1	\\
  	2FGLJ0158.3-3931 		& J015752.11-392906.2 		&                                					& 1.17(0.04) & 3.19(0.05) & 2.58(0.11) & BZQ 		& C & M   		& 		& 1	\\
  	2FGLJ0205.4+3211 		& J020537.46+321812.7 		&                                					& 0.96(0.04) & 2.58(0.06) & 2.38(0.16) & UND 		& C & M   		& 		& 1	\\
  	2FGLJ0217.5-0813 		& J021649.93-080551.8 		& SDSSJ021649.94-080551.8        		& 1.11(0.04) & 3.23(0.05) & 2.45(0.12) & BZQ 		& C & s   		& 		& 3	\\
  	2FGLJ0217.5-0813 		& J021716.20-082816.2 		& SDSSJ021716.21-082816.3        		& 1.19(0.03) & 2.68(0.03) & 2.35(0.04) & UND 		& C & M   		& 0.497	& 3	\\
  	2FGLJ0219.1-1725 		& J021906.91-173135.3 		&                                					& 1.20(0.05) & 2.98(0.09) & 2.37(0.25) & BZQ 		& C & -    		& 		& 2	\\
  	2FGLJ0222.0-1615 		& J022222.65-162309.6 		&                                					& 1.11(0.06) & 3.04(0.10) & 2.68(0.24) & BZQ 		& C & -    		& 		& 0	\\
	\tableline
	\label{tab:wgscatalogsamplealternative}
	\end{tabular}
		\begin{tablenotes}[para]
 Ê Ê Ê Ê Ê	\item {Notes:}\\
 Ê Ê Ê Ê Ê	\item[a] 2FLG name of the $\gamma$-ray source associated\\
		\item[b] \wse\ name of the candidate blazars\\
		\item[c] alternative name (if available) of the source in the literature\\
		\item[d] $c_{1}$ \wse\ color of the candidate\\
		\item[e] $c_{2}$ \wse\ color of the candidate\\
		\item[f] $c_{3}$ \wse\ color of the candidate\\
		\item[g] classification of the candidate blazar according to our method\\
		\item[h] class of the candidate blazar according to our method\\ 
		\item[i] note about the available multi-wavelength information of the \wse\ source, if already observed and/or classified 
		(surveys: N=NVSS, F=FIRST, S=SUMSS, M=2MASS, s=SDSS DR8, 6=6dFG, x=XMM or {\it Chandra}, X=ROSAT;  
		classification: QSO=quasar, Sy=Seyfert, LNR=LINER; variability: v=variable in \wse\ (var$\_\mathrm{flag} >$ 5 in at least one 
		\wse\ filter))\\
		\item[l] redshift of the sources\\
		\item[m] number of \wse\ sources in the background region (BR) of the $\gamma$-ray source selected as candidate 
		blazars with class equal or higher than the class of the best candidate blazar selected in the SR of the high-energy source.
	\end{tablenotes}
	\end{threeparttable}
	\end{center}
\end{table*}

\section{Efficiency and completeness}
\label{sec:efficomp}

The new parametrization of the \emph{locus}, coupled with the revisited 
association procedure described in Section~\ref{sec:association}, can be treated 
as a \emph{classifier} whose parameters are optimized through a supervised learning procedure. The training sample of this 
classifier is the WFB sample that has been used to define the 
\emph{locus} model in the PCs space generated by the WFB distribution in the \wse\ 
color space, because the association procedure is based on the geometry of the \emph{locus}. 
More specifically, the training of the classifier consists in the 
characterization of the \emph{locus} model that is used to calculate the scores 
and to select the candidate blazars. In general, after a classifier has been trained its 
performances can be evaluated by applying the trained classifier 
on a different sample of sources, the testset, representative of the same underlying 
population from where the training set has been extracted. 
The performance of a classification method can be expressed by two quantities, 
the efficiency and the completeness of the classification. The efficiency is the ratio between
the number of sources correctly classified by the classifier in the test-set and the total number 
of sources classified, and the completeness is the fraction of sources that the method correctly 
classifies relative to the total number of sources in the test-set that would have been correctly
classified by an ideal perfect classifier. 

The classification performed by the association procedure
on one $\gamma$-ray source of the WFB sample is considered correct if one
of the \wse\ sources associated with the WFB $\gamma$-ray source is the 
\wse\ source identified as counterpart of the WFB source by the method 
described in Section~\ref{sec:sample})\footnote{The candidate blazars that do not 
correspond to the \wse\ counterparts of the blazars in the WFB sample do not 
necessarily are incorrect from the physical point of view, as they can possibly represent
more physically meaningful associations of the high-energy source worth further investigation. We
label them as ``incorrect'' only to simplify the description of the 
evaluation of the efficiency and completeness of the association procedure.}. 
The efficiency $e$ and completeness $c$ are thus defined as:

\begin{eqnarray}
	e\!=\!\frac{n(\mathrm{Correctly\ associated\ } \gamma\mathrm{-ray\ sources})}
	        {n(\mathrm{Associated\ } \gamma\mathrm{-ray\ sources)}}\nonumber\\
	c\!=\!\frac{n(\mathrm{Correctly\ associated\ } \gamma\mathrm{-ray\ sources})}
	        {n(\gamma\mathrm{-ray\ sources\ to\ be\ associated)}}
	\nonumber	        
\end{eqnarray}

When the size of the parent population of the sample of sources used to train and test the
classifier is large enough, training set and testset can be obtained by splitting the parent
population in two subsets of different sizes. The minimum size of the sample that would 
permit to apply this strategy depends on the specific problem, on the features of the 
classifier used and the sample itself. Common choices of the training set to testset 
size proportions are 60\%-40\% or 70\%-30\%. Using this approach to assess the 
performances of our classifier, we verified that the values of the efficiency and 
completeness of the association strongly depended
on the composition of both samples, revealing a suboptimal training of the classifier 
that led to ``over-fitting''.
Two likely causes of this behavior are, respectively, that our classifier is too complex to be trained on 
60\% or 70\% fraction of the WFB sample, and that the parameters of the \emph{locus} model are
very sensitive to the particular subset of sources contained in the training test. To avoid
this issue, we have used all WFB sources to determine the \emph{locus} modelization, as
described in Section~\ref{sec:locus}, and estimated 
the efficiency $e$ and completeness $c$ of the association 
procedure using
a different strategy, the $K$-fold cross-validation \cite{hastie2009}, that employs the same sample used
as training set. With the $K$-fold cross-validation 
approach, the WFB sample is randomly 
partitioned into $K$ equal sized subsamples. Of the $K$ subsamples, only one subsample 
is retained as the testset data to evaluate the performances of the association procedure, 
while the remaining $K-1$ subsamples are used as training data of the \emph{locus} model. 
Thus the efficiency and completeness are evaluated on the $K$-th subset not used for 
training. The same steps are then repeated $K$ times, with each of the $K$ subsets of the
WFB sample used exactly once as testset. The efficiency and completeness values determined
for each of the $K$ cross-validations can be combined to produce single robust estimates.

\begin{figure*}[]
	\includegraphics[height=9cm,width=9cm,angle=0]{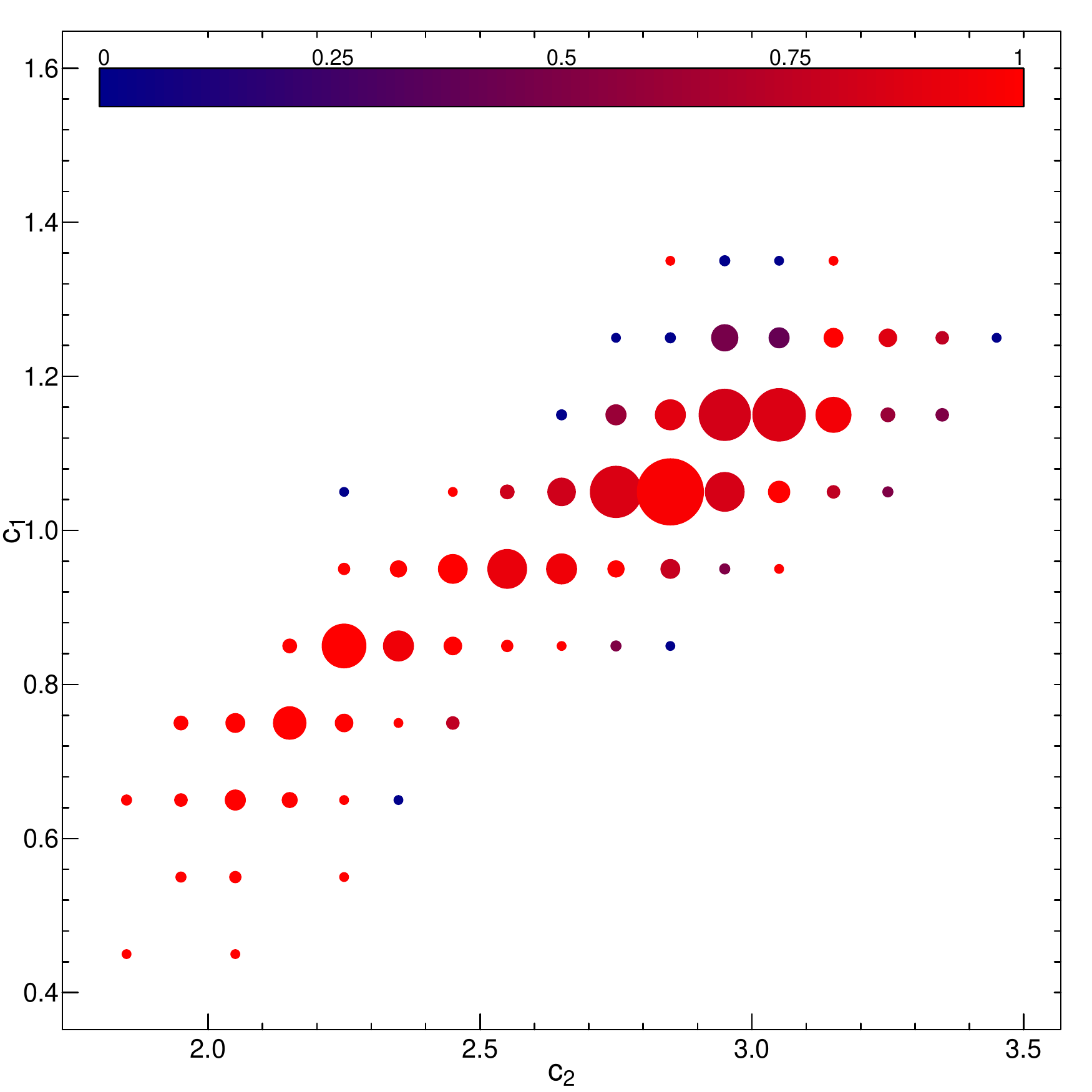}
        	\includegraphics[height=9cm,width=9cm,angle=0]{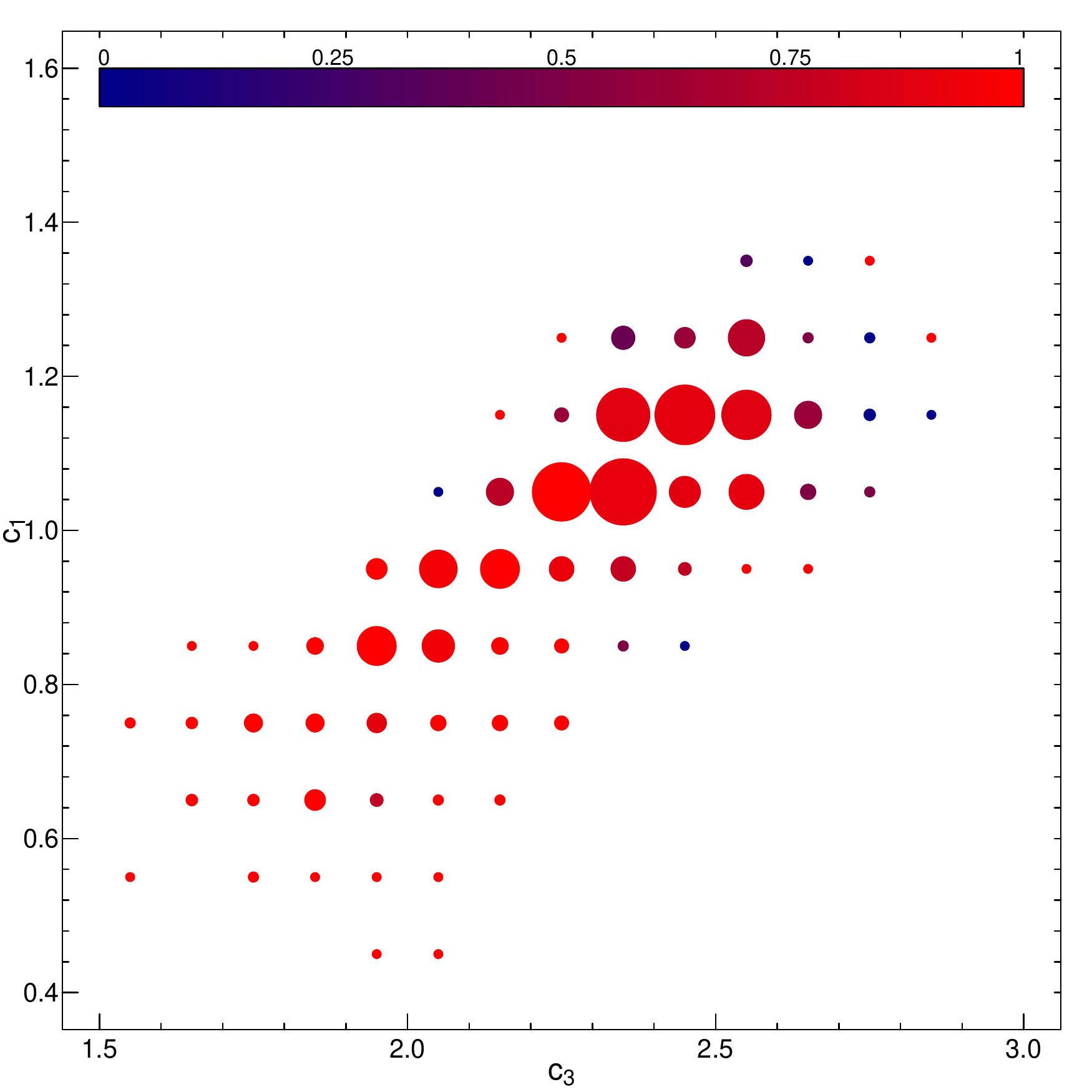}\\
        	\includegraphics[height=9cm,width=9cm,angle=0]{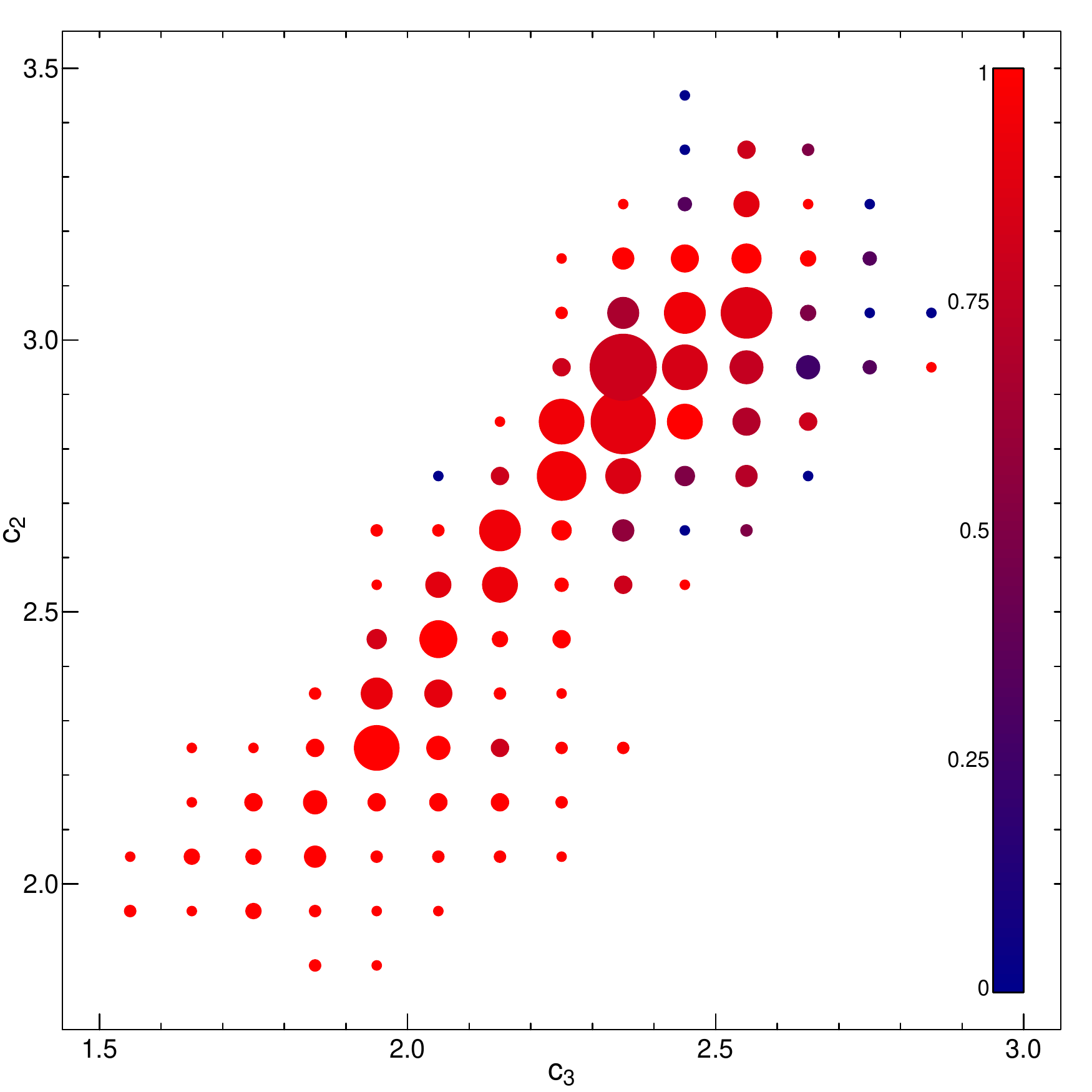}
        	\includegraphics[height=9cm,width=9cm,angle=0]{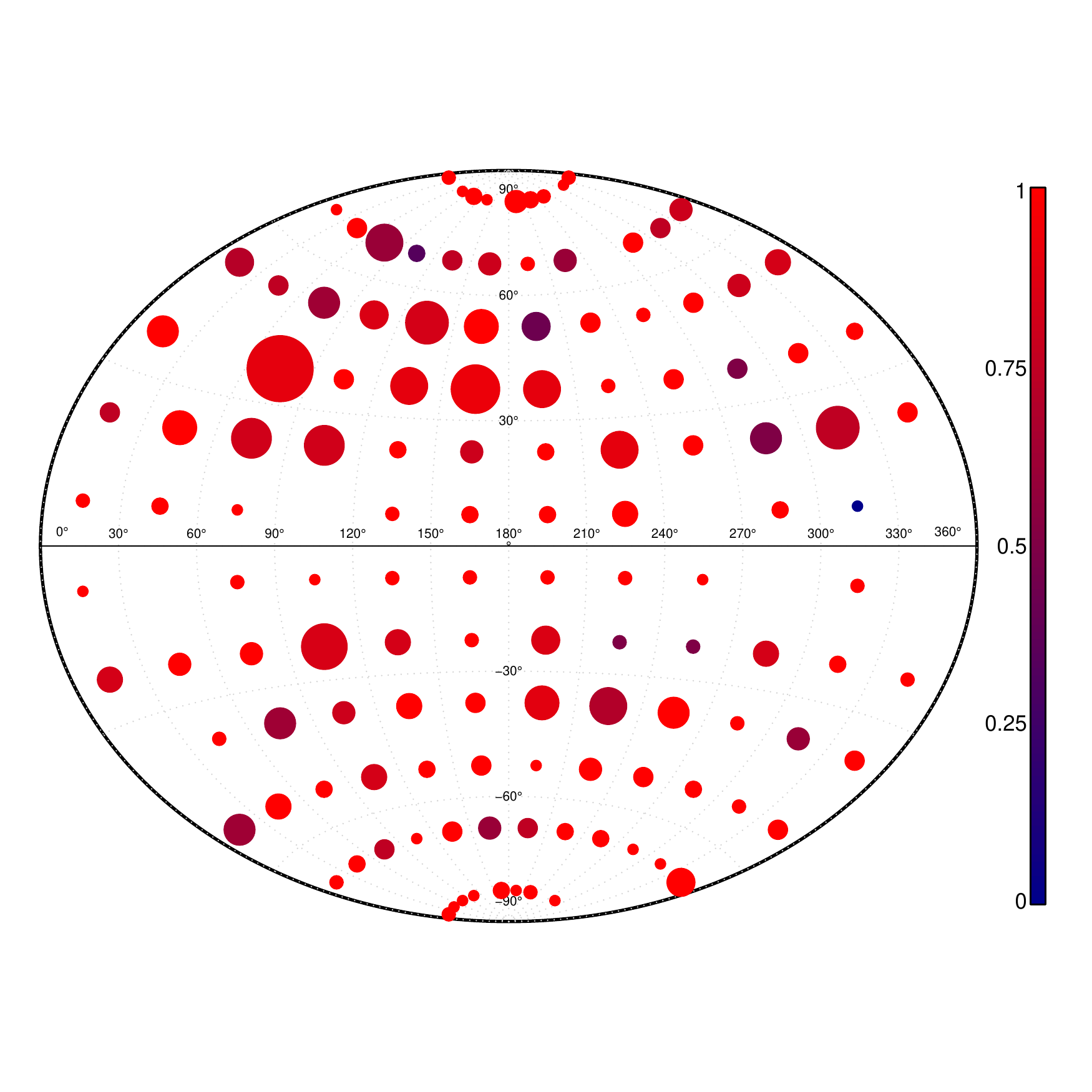}           
          \caption{Maps of the efficiency $e$ of the association procedure based on the new
          model of the WFB \emph{locus} in the \wse\ color space obtained with the 100-fold
          cross-validation as described in Sec.~\ref{sec:efficomp}. The upper-left plot shows 
          $e$ in the $[3.4]-[4.6]$ vs $[4.6]-[12]$
          color-color plane, the upper-right plot shows $e$ in the $[4.6]-[12]$ vs $[12]-[22]$
          color-color plane and the lower-right plot shows $e$ in the $[3.4]-[4.6]$ vs $[12]-[22]$ 
          color-color plane. The lower-right plot shows the efficiency as a function of the galactic
          coordinates of the $\gamma$-ray sources in the WFB sample. In all plots, the size of the symbols
          is proportional to the number of WFB sources in the each bin and the local value of the 
          efficiency normalized to unity is color-coded.}
          \label{fig:efficiencymaps}
\end{figure*}

\begin{figure*}[] 
	\includegraphics[height=9cm,width=9cm,angle=0]{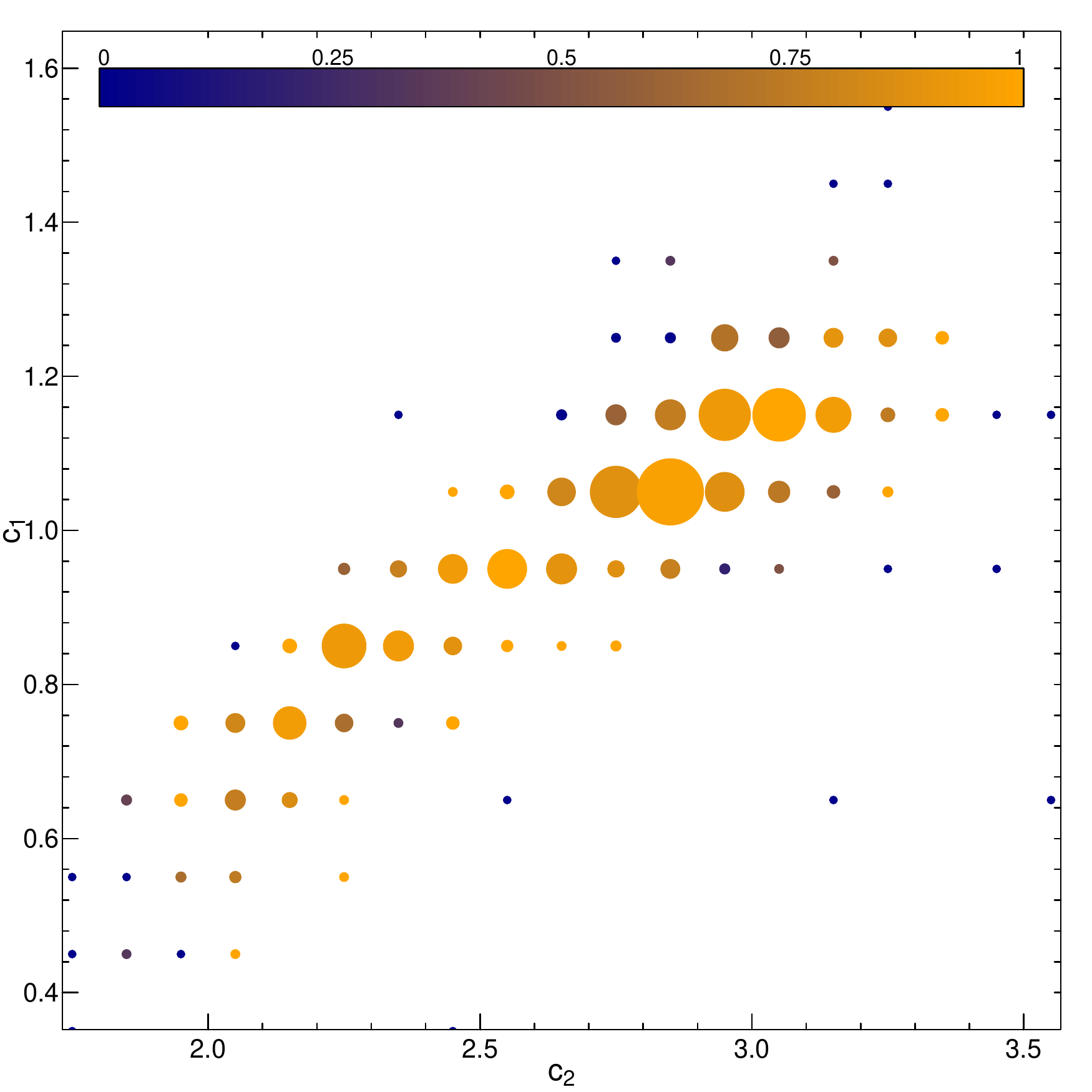}
        	\includegraphics[height=9cm,width=9cm,angle=0]{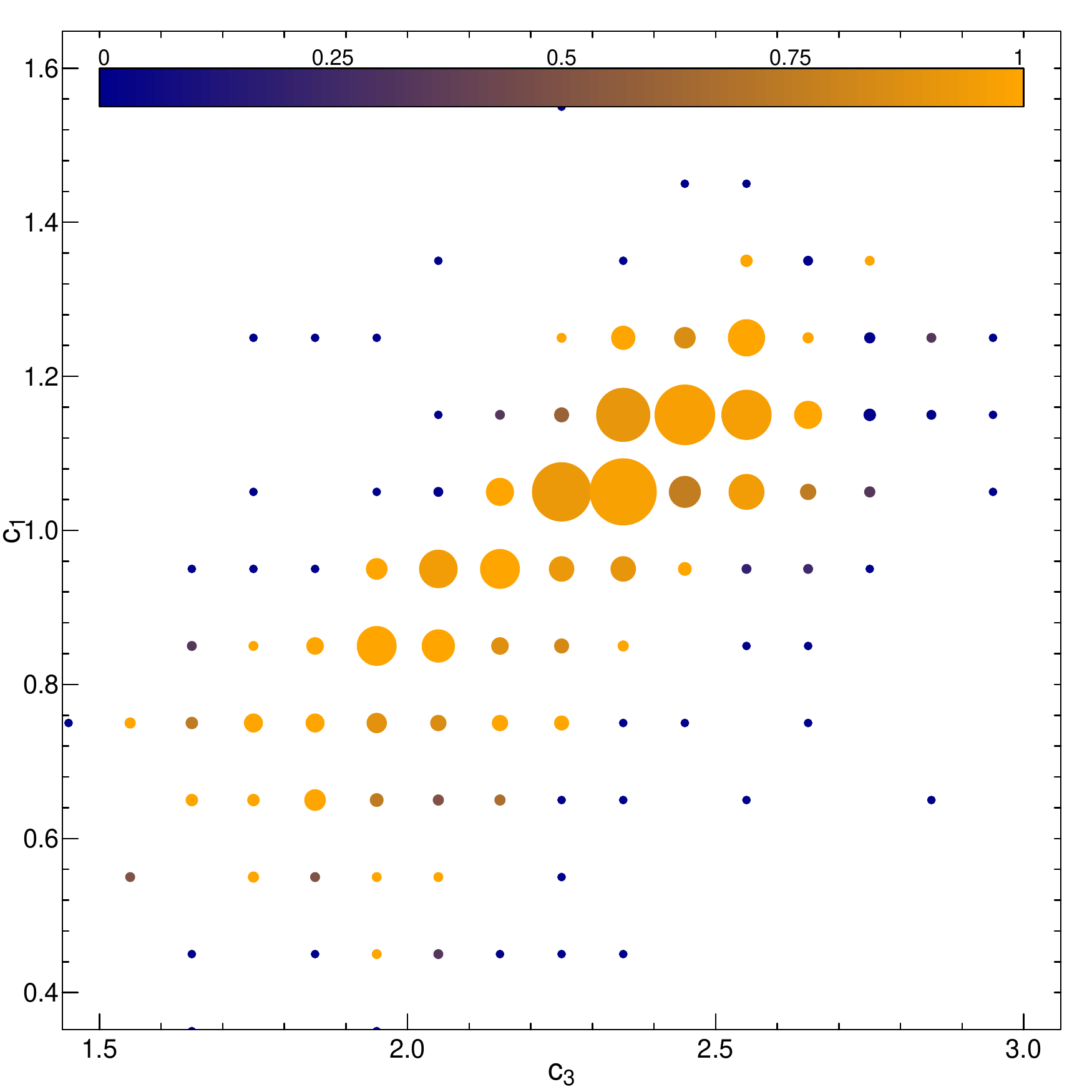}\\
        	\includegraphics[height=9cm,width=9cm,angle=0]{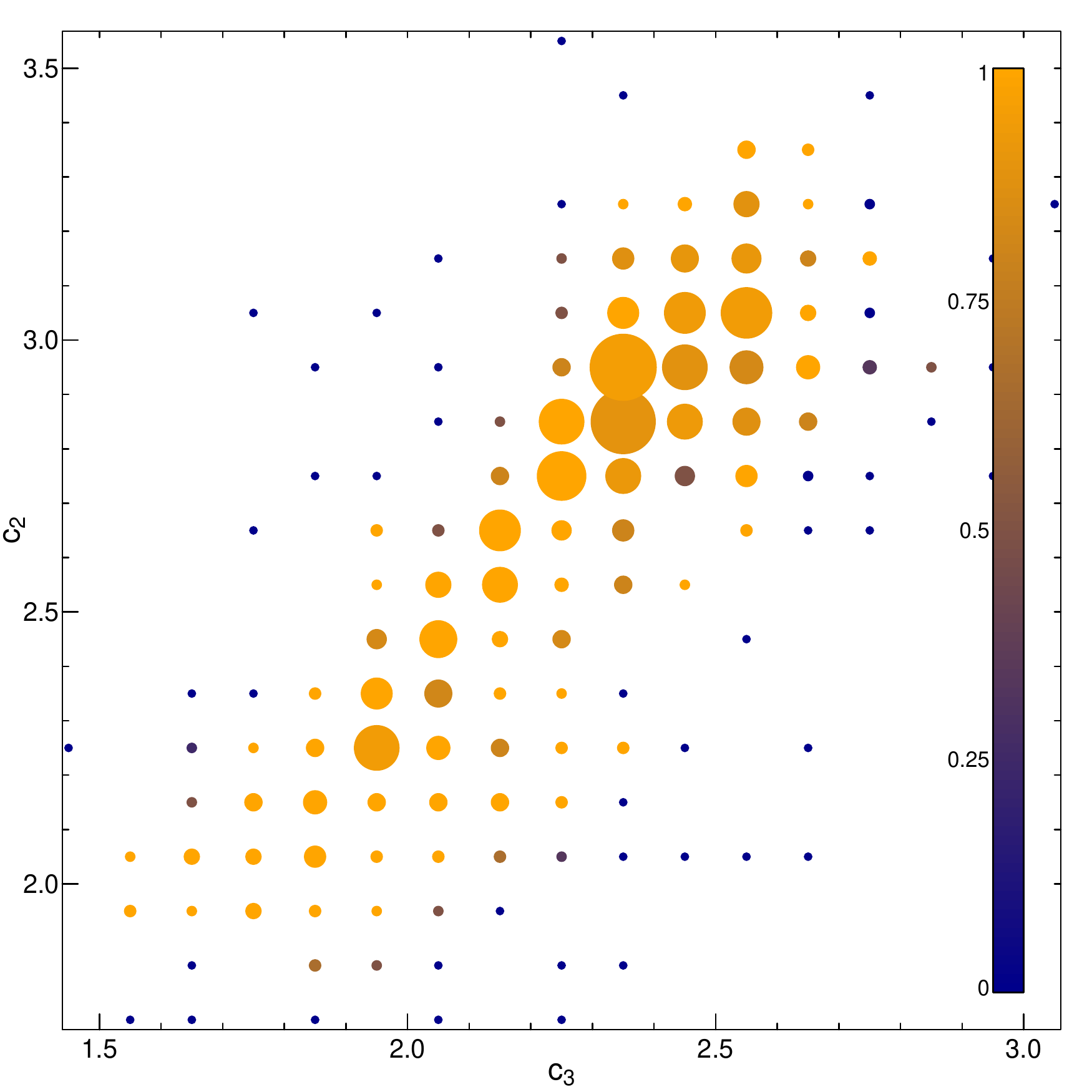}
         	\includegraphics[height=9cm,width=9cm,angle=0]{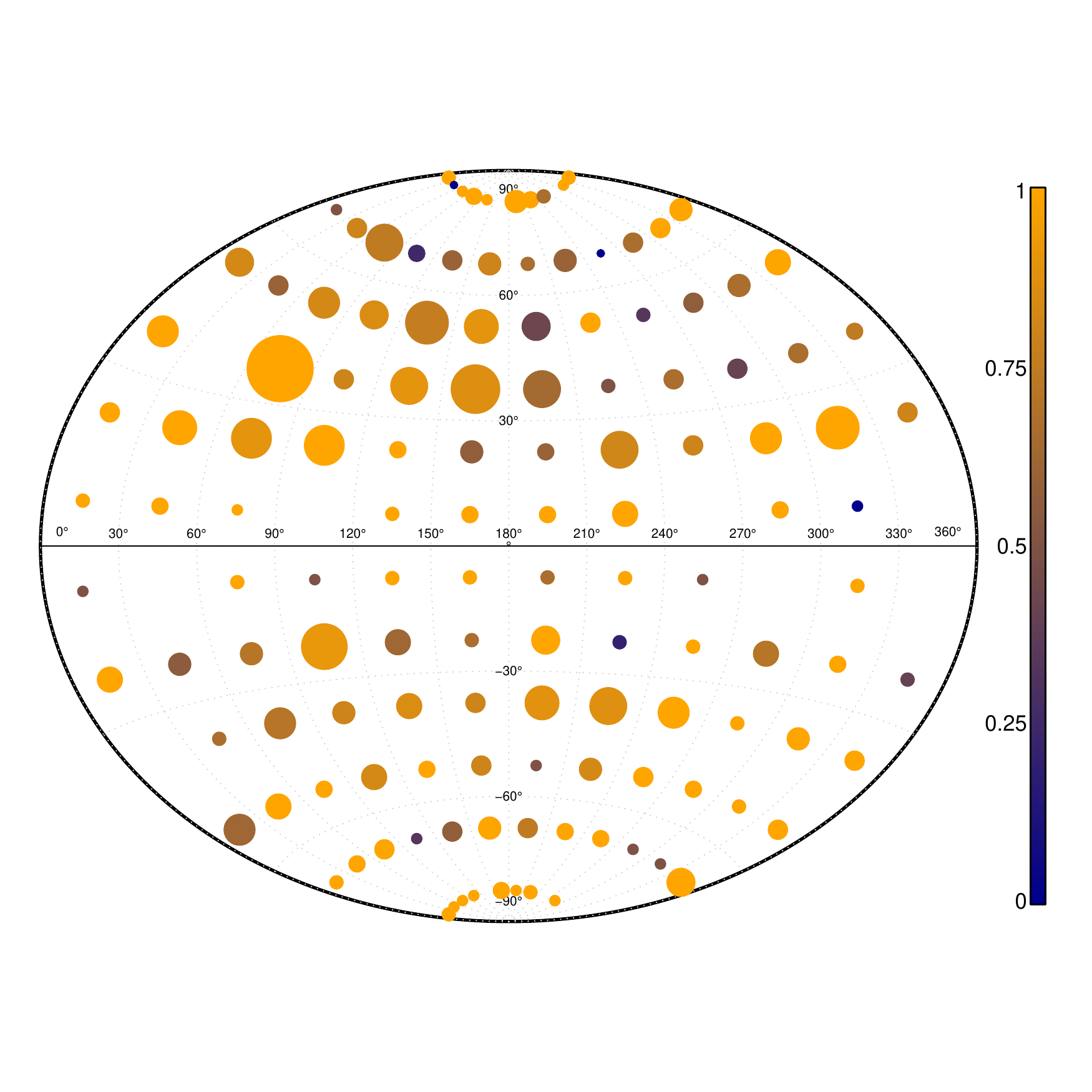}           
	\caption{Maps of the completeness $c$ of the association procedure based on the new
          model of the WFB \emph{locus} in the \wse\ color space obtained with the 100-fold
          cross-validation as described in Sec.~\ref{sec:efficomp}. The upper-left plot shows $c$ 
          in the $[3.4]-[4.6]$ vs $[4.6]-[12]$
          color-color plane, the upper-right plot shows $c$ in the $[4.6]-[12]$ vs $[12]-[22]$
          color-color plane and the lower-right plot shows $c$ in the $[3.4]-[4.6]$ vs $[12]-[22]$ 
          color-color plane. The lower-right plot shows the completeness as a function of the galactic
          coordinates of the $\gamma$-ray sources in the WFB sample. In all plots, the size of the symbols
          is proportional to the number of WFB sources in the each bin and the local value of the 
          efficiency normalized to unity is color-coded.}
       	\label{fig:completenessmaps}
\end{figure*}

\noindent The parameters of the $K$ different models of the \emph{locus} estimated with the 
cross-validation on the sum of the $K$-1 subsets not used for validation are compatible with the 
values obtained with the procedure described in Section~\ref{subsec:model} for $K\!=\!\{20,30,50,100\}$. 
The total efficiency and completeness
for the re-association of the WFB sources performed with the new association 
procedure are $e_{\mathrm{tot}}\!\simeq\!97\%$ and $c_{\mathrm{tot}}\!\simeq\!81\%$ respectively. 
Both $e$ and $c$ have been also estimated as functions 
of the \wse\ colors and the galactic coordinates of the \wse\ sources (Figure~\ref{fig:efficiencymaps} for 
the efficiency maps and Figure~\ref{fig:completenessmaps} for the completeness maps). 
The maps of $e$ and $c$ in the color-color planes and galactic coordinates have been created by 
estimating the two quantities in bins of the three independent \wse\ color-color projections and 
galactic coordinates respectively. For example, the efficiency $e^{(c_{1}c_{2})}_{ij}$ and completeness
$c^{(c_{1}c_{2})}_{ij}$ evaluated in the $ij$-th bin of the first \wse\ color-color plane 
are defined as:

\begin{eqnarray}
	e^{(c_{1}c_{2})}_{ij}\!=\!e(c_{1}(i)\!\leq\!c_{1} \!<\!c_{1}(i+1),c_{2}(i)\!\leq\!c_{2}\!<\!c_{2}(i+1))\nonumber\\
	c^{(c_{1}c_{2})}_{ij}\!=\!c(c_{1}(i)\!\leq\!c_{1} \!<\!c_{1}(i+1),c_{2}(i)\!\leq\!c_{2}\!<\!c_{2}(i+1))\nonumber
\end{eqnarray}

Similarly, we evaluate $e$ and $c$ in the $ij$-th bin of the galactic coordinates 
distribution of the WFB \wse\ counterparts as follows:

\begin{eqnarray}
	e^{(lb)}_{ij}\!=\!e(l_{1}(i)\!\leq\!l \!<\!l(i+1), b(i)\!\leq\!b\!<\!b(i+1))\nonumber\\
	c^{(lb)}_{ij}\!=\!c(l_{1}(i)\!\leq\!l \!<\!l(i+1), b(i)\!\leq\!b\!<\!b(i+1))\nonumber
\end{eqnarray}

The plots in Figure~\ref{fig:efficiencymaps} and Figure~\ref{fig:completenessmaps} show 
the values of the normalized efficiency and completeness expressed as the color of the 
symbols, while the size of the symbols are proportional to the total number of $\gamma$-ray 
candidate blazars falling in each bin of the maps. The right lower
plots in both figures show the values of $e$ and $c$ as functions of the galactic coordinates in 
Aitoff projection. One comment that can be made by observing the efficiency map in galactic 
coordinates in the lower-right plot in Figure~\ref{fig:efficiencymaps} is that the contribution of galactic 
sources to the contamination of the association procedure is not dominant  as there is no clear 
indication of a positive gradient of the efficiency as a function of the galactic latitude (namely, going
from the galactic disk to the galactic north and south poles).

\section{Association of the Fermi $gamma$-ray blazars}
\label{sec:association2fb}

\begin{table*}
	\tiny
	\begin{center}
	\begin{threeparttable}
	\caption{First ten candidate blazars associated to 2FB sources (the complete list of 2FB associations can be found in 
	Table~\ref{tab:2fbcatalogreassociated} in Appendix B).}
	\begin{tabular}{lllccccclclc}
	\tableline\tableline
	2FGL\tnote{a}  &  \wse\tnote{b}	&  other\tnote{c} 	& [3.4]-[4.6]\tnote{d} 	& [4.6]-[12]\tnote{e} 	& [12]-[22]\tnote{f} 	& type\tnote{g} 	& class\tnote{h}& notes\tnote{i} & z\tnote{l}	& reassoc.\tnote{m}  & $N_{\mathrm{BR}}$\tnote{n}	\\
  	name 	        &  name   		&  name  			&     mag     		&    mag     		&   mag     			&      			&       		&       		&   			& flag   			& 						\\
	\tableline
	2FGLJ0035.8+5951 		&    J003552.62+595004.3 	&    BZBJ0035+5950           	&    0.59(0.03)&   2.03(0.03) &   1.87(0.09) &     BZB   &     B     &     M,v (0.08)6? &     y     &     0    \\	
	2FGLJ0047.2+5657 		&    J004700.43+565742.4 	&    BZUJ0047+5657           	&   1.03(0.04) &   2.56(0.06) &   2.4(0.13)  	 &     UND   &     C     &     v     &     0.747  &     y     &     1    \\  	
  	2FGLJ0057.9+3311 		&    J005832.05+331117.3 	&    BZUJ0058+3311           	&    1.13(0.05) &  3.0 (0.07)  &     2.45(0.19)  &     BZQ   &     C     &     -     &     1.369  &     y     &     4    \\
  	2FGLJ0102.7+5827 		&    J010245.75+582411.1 	&    BZUJ0102+5824           	&    1.06(0.03) &  2.98(0.03)  &     2.45(0.05)  &     BZQ   &     C     &     v     &     0.644? &     y     &     0    \\
  	2FGLJ0105.3+3930 		&    J010509.20+392815.2 	&    BZUJ0105+3928          	 &    1.01(0.03) &     2.5(0.03)  &     2.16(0.05)  &     UND   &     C     &     v   (0.08)3? &     y     &     1    \\
  	2FGLJ0105.3+3930 		&    J010542.74+393024.2 	&    -                       			&    1.17(0.06) &     3.14(0.1)   &     2.55(0.26)  &     BZQ   &     C     &     -     &     ?      &     n     &     -    \\
	2FGLJ0109.9+6132 		&    J010946.32+613330.4 	&    NVSSJ010946+613329      &    1.06(0.03) &     2.71(0.03)  &     2.47(0.05)  &     UND   &     B     &     v     &     0.783  &     y     &     0\\ 
	2FGLJ0113.2-3557 		&    J011315.83-355148.2 	&    BZQJ0113-3551           	&    1.12(0.04) &     2.9(0.05)  &     2.48(0.1)  &     BZQ   &     C     &     -     &     1.22   &     y     &     2    \\
	2FGLJ0114.7+1326 		&    J011452.77+132537.6 	&    BZBJ0114+1325           	&    0.82(0.03) &     2.3(0.04)  &     1.86(0.15)  &     BZB   &     B     &     M     &     ?      &     y     &     0    \\
	2FGLJ0144.6+2704 		&    J014433.54+270503.1 	&    BZUJ0144+2705           	&    1.07(0.03) &     2.75(0.03)  &     2.22(0.04)  &     UND   &     B     &     M,v   &     ?      &     y     &     0    \\
	\tableline
	\label{tab:catalogsample2fb}
	\end{tabular}
	\begin{tablenotes}[para]
 Ê Ê Ê Ê Ê	\item {Notes:}\\
 Ê Ê Ê Ê Ê	\item[a] 2FLG name of the $\gamma$-ray source associated\\
		\item[b] \wse\ name of the candidate blazars\\
		\item[c] alternative name (if available) of the source in the literature\\
		\item[d] $c_{1}$ \wse\ color of the candidate\\
		\item[e] $c_{2}$ \wse\ color of the candidate\\
		\item[f] $c_{3}$ \wse\ color of the candidate\\
		\item[g] classification of the candidate blazar according to our method\\
		\item[h] class of the candidate blazar according to our method\\ 
		\item[i] note about the available multi-wavelength information of the \wse\ source, if already observed and/or classified 
		(surveys: N=NVSS, F=FIRST, S=SUMSS, M=2MASS, s=SDSS DR8, 6=6dFG, x=XMM or {\it Chandra}, X=ROSAT;  
		classification: QSO=quasar, Sy=Seyfert, LNR=LINER; variability: v=variable in \wse\ (var$\_\mathrm{flag} >$ 5 in at least one 
		\wse\ filter))\\
		\item[l] redshift of the sources\\
		\item[m] flag indicating whether our association corresponds to the association provided in the 2FGL catalog (``y''), 
		or otherwise (``n'')\\
		\item[n] number of \wse\ sources in the background region (BR) of the $\gamma$-ray source selected as candidate 
		blazars with class equal or higher than the class of the best candidate blazar selected in the SR of the high-energy source.
	\end{tablenotes}
	\end{threeparttable}
	\end{center}
\end{table*}

The sample of $\gamma$-ray blazars associated by the 2FGL contains 752 $\gamma$-ray sources,
excluding all sources with any analysis flag. We excluded from this list the 610 blazars already 
contained in WFB sample, for a final number of 142 2FGL sources. These $\gamma$-ray sources 
constitute the 2FGL \fer\ blazar sample (hereinafter 2FB). The 2FB sample has been investigated 
with our association procedure similarly at what done for the WFB sample. The \emph{locus} model 
and the parameters of the association procedure applied to the 2FB are the same used for the 
re-association of the WFB sample. The first ten associations of the 2FB sources are shown in 
Table~\ref{tab:catalogsample2fb}, while the summary of the composition of the sample of 
candidate blazars associated with the 2FB sources can be found in Table~\ref{tab:composition}. 
We summarize in the following the number and type of spurious candidate blazars found in the 
BRs of the 2FB sources.
No class A spurious candidate blazars have been found within the 
BRs of the 8 2FB sources associated to at least one class A candidate blazar (0\%), 
only 1 out of the 34 2FB sources associated to a class B or better candidate
has one class B candidate blazar in the BR ($\sim\!3\%$). The number of 2FB sources associated
to class C candidate blazars with at least one candidate blazars of same or better class is 31 out of the 
remaining 34 sources, corresponding to the $\sim\!91\%$ of this sample.
Also in this case, an extensive archival research has been performed for all candidate 
blazars associated by our method. This research has led to gather the essential information 
about the candidate blazars contained in the tenth column of Table~\ref{tab:catalogsample2fb}.

\section{Summary and conclusions}
\label{sec:summary}

Using the preliminary data release of the Wide-field Infrared Survey Explorer (\wse), we 
discovered that $\gamma$-ray emitting blazars have infrared colors that distinguish them 
from other galactic and extragalactic sources in the 3-dimensional IR color space (Papers I and II).  
Then, we used these results to develop an association method able to associate IR selected 
blazar candidates as low-energy counterparts of a $\gamma$-ray source (Papers III and IV).

In this paper we have described an updated version of the WFB sample, gathered using the new \wse\
All-Sky release, the 2FGL catalog and the latest release of the \bzcat\ list of blazars.
Then, we have discussed a new association procedure for the unidentified high-energy sources
based on a new model of the \emph{locus} occupied by WFB sample in the three-dimensional
PCs space generated by the distribution of WFB \wse\ sources in the \wse\ color space. We defined
a quantitative measure of the compatibility of a generic \wse\ source with the \emph{locus} model
and expounded the new association procedure. The new association procedure can 
select candidate blazars classified as BZB or BZQ candidates, and ranked according to the likelihood
of each candidate of being an actual blazar. We also investigated the possibility of spurious 
associations by determining the number and class of \wse\ sources compatible with the model of the 
WFB \emph{locus} in background regions defined around the SR of each high-energy source.
We have assessed the performances of the association procedure in terms of the efficiency 
and completeness by re-associating the $\gamma$-ray sources in the WFB sample, yielding a 
total efficiency $e_{\mathrm{tot}}\!\simeq\!97\%$ and total completeness $c_{\mathrm{tot}}\!\simeq\!81\%$ 
respectively. By using a $K$-fold cross-validation approach, we have also estimated the efficiency and 
completeness as functions of the \wse\ colors and galactic coordinates of the candidate blazars.
The lack of a positive gradient in the efficiency of the association procedure as function of the 
galactic latitude suggests that galactic sources do not contribute significantly to the 
contamination of the \emph{locus}. We will make the code for the association procedure 
available to the astronomical community on request. 

In this paper, we have presented the catalog of candidate blazars associated by the new 
procedure to the 2FGL $\gamma$-ray sources included in the WFB sample,
used to define the new model of the \emph{locus}. We have investigated the archival 
information available for the \wse\ sources
representing alternative associations of the \fer\ $\gamma$-ray sources in the WFB.
We also presented the catalog of 
candidate blazars obtained by applying the new association procedure to the 2FB sample, 
composed of all clean $\gamma$-ray sources associated with blazars in the 2FGL catalog 
but not contained in the WFB sample. In 
both catalogs, for every candidate blazar we provided the basic \wse\ data and information about the 
association; for the alternative associations (candidate blazars different from the \wse\ counterparts
defined in the WFB and 2FB samples), we complemented the \wse\ data with some additional archival
information (known name and classification), when available, that can hopefully help to physically 
characterize the nature of these sources. We will make both catalogs publicly available 
in electronic format.

Finally, we acknowledge that the effects of the distribution in redshift and the variability in \wse\ 
observations of the WFB sources on the definition of the model of the \emph{locus} on which the 
association procedure is based are still unknown. To address these open questions, we plan to carry out 
a detailed investigation of the \wse\ blazar variability and the statistical characterization of the SEDs of 
WFB blazars, that will be presented in future papers.

\acknowledgements

The work is supported by the NASA grants NNX10AD50G, NNH09ZDA001N and 
NNX10AD68G. R. D'Abrusco gratefully acknowledges the financial 
support of the US Virtual Astronomical Observatory, which is sponsored by the
National Science Foundation and the National Aeronautics and Space Administration.
F. Massaro is grateful to A. Cavaliere, S. Digel, D. Harris, D. Thompson, A. Wehrle for 
their helpful discussions.
The work by G. Tosti is supported by the ASI/INAF contract I/005/12/0.
H. A. Smith acknowledges partial support from NASA/JPL grant RSA 1369566.
TOPCAT\footnote{\underline{http://www.star.bris.ac.uk/$\sim$mbt/topcat/}} 
\citep{taylor2005} and SAOImage DS9 were used extensively in this work 
for the preparation and manipulation of the tabular data and the images.
This research has made use of data obtained from the High Energy 
Astrophysics Science Archive
Research Center (HEASARC) provided by NASA's Goddard
Space Flight Center; the SIMBAD database operated at CDS,
Strasbourg, France; the NASA/IPAC Extragalactic Database
(NED) operated by the Jet Propulsion Laboratory, California
Institute of Technology, under contract with the National Aeronautics and Space Administration.
Part of this work is based on archival data, software or on-line services provided by the ASI Science Data Center.
This publication makes use of data products from the Wide-field Infrared Survey Explorer, 
which is a joint project of the University of California, Los Angeles, and 
the Jet Propulsion Laboratory/California Institute of Technology, 
funded by the National Aeronautics and Space Administration.

{}

\newpage
\section{Appendix A: performances of the association procedure as functions of index of the score assignment law}
\label{app:phivariations}

\begin{figure}[] 
	\begin{center}
	\includegraphics[height=8.5cm,width=8.5cm,angle=0]{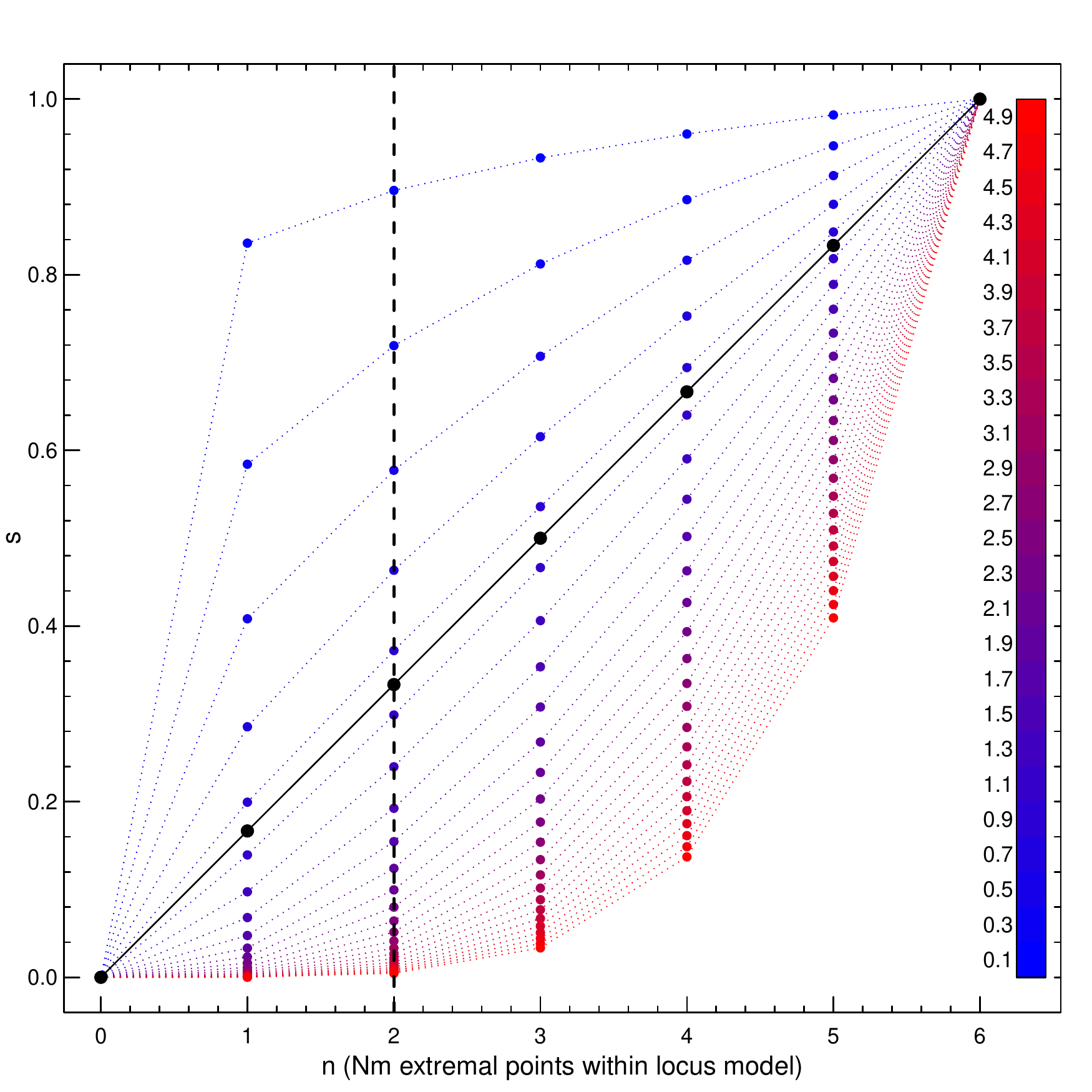}
	\includegraphics[height=8.5cm,width=8.5cm,angle=0]{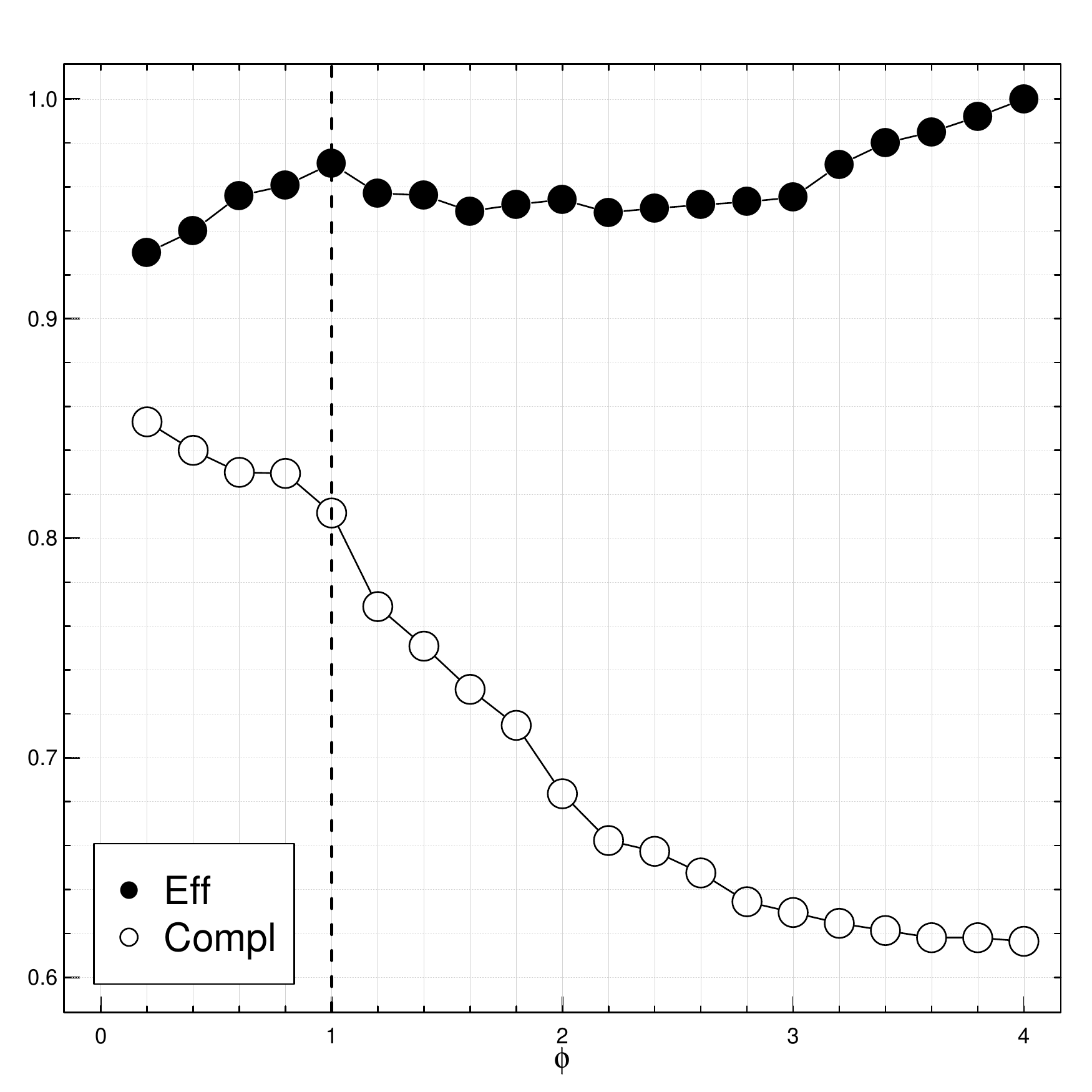}
          \caption{Left panel: score as function of the number of extremal points $n$ contained in 
          the \emph{locus} model for different values of the parameter $\phi$ of the score assignment law
          (Equation~\ref{eq:score}). For a fixed number 
          of extremal points $n$ within the model, the corresponding value of the score will be larger
          than $n$/6 for $\phi<1$ and smaller than $n$/6 for $\phi<1$ (the dashed vertical black line
          shows the score values for $n=2$ extremal points as an example). The linearly proportional 
          assignment law for $\phi=1$ 
          is represented by the curve with black solid circles. Right panel: Total efficiency $e_{\mathrm{tot}}$ 
          (solid circles) and completeness $c_{\mathrm{tot}}$
          (open circles) of the association procedure evaluated on the whole WFB sample as functions of the 
          parameter $\phi$ of the score assignment law (Equation~\ref{eq:score} in Section~\ref{subsec:score}).
          The values of $e_{\mathrm{tot}}$ and $c_{\mathrm{tot}}$ obtained with $\phi\!=\!1$, used for 
          the association of the 
          WFB and 2FB samples in Section~\ref{sec:associationwfb} and Section~\ref{sec:association2fb} 
          respectively, are indicated by the vertical dashed line.}
          \label{fig:phivariation}
	\end{center}          
\end{figure}

The score assignment law (Equation~\ref{eq:score}) has been introduced in Section~\ref{subsec:score} to 
provide a flexible way to assign the score values to the \wse\ sources as a function of the number
of extremal points contained within the \emph{locus} model. As shown in the left panel of 
Figure~\ref{fig:phivariation}, the score of a source with $n$ extremal points contained in one
cylinder of the \emph{locus} model for $\phi \!>\!1$ becomes smaller than the fractional value of 1/$n$, 
while for $\phi<1$ the score becomes larger than the default value corresponding to the linear 
proportionality obtained with $\phi=1$. The parameter $\phi$ permits to control the behavior of the 
score assignment law and to tweak it for different scientific goals. For example, a more efficient 
(or pure) association can be obtained by assigning large score values only to the \wse\ sources 
with a large number of extremal points 
within the model; this can be accomplished by using a value of the index larger than unity. On the 
other hand, by assigning large
scores to sources with few extremal points inside the model, the completeness of the selection can be 
enhanced at the cost of a lower efficiency. For this reason, the value of $\phi$ has to be optimized for each 
distinct experiment. In the case of the experiments described in this paper,
the choice of the value of the parameter $\phi$ has been based on the characterization of the efficiency 
and completeness (defined in Section~\ref{sec:efficomp}) of the association procedure 
as functions of $\phi$.
We have run multiple association experiments on the WFB sample with fixed values of the
parameters of the geometric model of the \emph{locus} (see Table~\ref{tab:model}), 
but letting $\phi$ vary in the $(0.2, 4)$ range. Values of $e_{\mathrm{tot}}$ and $c_{\mathrm{tot}}$
have been evaluated for these experiments according to the definitions given 
in Section~\ref{sec:efficomp}. The distributions of the $e_{\mathrm{tot}}$ and $c_{\mathrm{tot}}$ 
as functions of $\phi$ are shown in the right panel in Figure~\ref{fig:phivariation}.

\noindent The curve representing $e_{\mathrm{tot}}$ varies in the range $(0.92,1)$ but 
is not monotonically increasing with $\phi$. The total efficiency locally peaks at $\sim\!0.97$
for $\phi\!=\!1$ and then decreases to $\sim\!0.95$ for larger values of $\phi$. 
The plateau observed between $\phi\!=\!1.4$ and 
$\phi\!=\!1.3$ indicates that a large fraction of candidate blazars have a large number of
extremal points within the cylindrical model of the \emph{locus}, so that the slowly rising curve in the
left panel in Figure~\ref{fig:phivariation} for small values of $n$ and the steeper slope 
for $n\!\geq\!4$ do not affect the overall efficiency of the association. The efficiency becomes 
larger and then reaches 1 for $\phi$ approaching 4. The total completeness 
$c_{\mathrm{tot}}$ is monotonically decreasing from small to large values of $\phi$, ranging
from a maximum value $\sim0\!.85$ for $\phi\!=\!0.1$ to a minimum value of $\sim0\!.62$ for
$\phi\!=\!4$. 
In general, the value of $\phi\!=\!1$ used for the evaluation of the efficiency
and completeness as functions of the color and galactic coordinates in Section~\ref{sec:efficomp}
and the association of the WFB and 2FB samples has been chosen as a reasonable 
compromise between the largest achievable total efficiency and a large 
enough completeness for the association procedure applied on the WFB sample. Different 
experiments should be performed with values of the parameter $\phi$ optimized for 
one aspect or the other of the association procedure.

\newpage
\section*{Appendix B: complete lists of \wse\ associations for WFB and 2FB samples}
\label{app:associations}

\begin{center}
\begin{longtable}{lllccccccl}
	\caption{List of all candidate blazars associated to the WFB sources. The explanation of the content of the columns can 
	be found in the notes of Table~\ref{tab:wfbcatalogsamplereassociated}.}\\
	\hline
	\hline
 	2FGL  	&  \wse\  	& \bzcat\ 	& [3.4]-[4.6] 	& [4.6]-[12] 	& [12]-[22] 	& z 	& 	type 	& class	&	$N_{\mathrm{BR}}$\tnote{m}	 	\\
  	name  	&  name   &  name   	&     mag     	&    mag     	&   mag     		&   	&      		&    		&								\\
	\hline
	\endfirsthead
	\multicolumn{10}{c}
	{\tablename\ \thetable\ -- \textit{Continued from previous page}} \\
	\hline
  	2FGL  	&  \wse\  	& \bzcat\ 	& [3.4]-[4.6] 	& [4.6]-[12] 	& [12]-[22] 	& z 	& 	type & class  	&	$N_{\mathrm{BR}}$\tnote{m}	 	\\
  	name  	&  name   &  name   	&     mag     	&    mag     	&   mag     		&   	&      		&    		&								\\
	\hline
	\endhead	
	\hline 
	\multicolumn{10}{r}
	{\textit{Continued on next page}} \\
	\endfoot
	\hline
	\endlastfoot
  2FGLJ0000.9-0748 & J000118.01-074626.7 & BZBJ0001-0746 & 0.93(0.03)& 2.53(0.04) & 2.12(0.1)& ?       & BZB & B & 0\\
  2FGLJ0004.7-4736 & J000435.64-473619.5 & BZQJ0004-4736 & 1.09(0.03)& 2.92(0.03) & 2.3 (0.06) & 0.88    & UND & C & 0\\
  2FGLJ0006.1+3821 & J000557.17+382015.2 & BZQJ0005+3820 & 1.1 (0.03)& 3.11(0.03) & 2.69(0.03) & 0.229   & BZQ & A & 0\\
  2FGLJ0007.8+4713 & J000759.97+471207.7 & BZBJ0007+4712 & 0.92(0.04)& 2.28(0.06) & 2.13(0.18)& 0.28    & BZB & C & 0\\
  2FGLJ0012.9-3954 & J001259.88-395425.8 & BZBJ0012-3954 & 1.01(0.04)& 2.84(0.04) & 2.28(0.11) & ?       & UND & C & 1\\
  2FGLJ0013.8+1907 & J001356.36+191042.0 & BZBJ0013+1910 & 0.97(0.04)& 2.58(0.06) & 2.09(0.22)& ?       & BZB & C & 1\\
  2FGLJ0017.4-0018 & J001611.08-001512.3 & BZQJ0016-0015 & 1.29(0.11)& 3.01(0.26)& 3.65(0.33) & 1.577   & -   & - & -\\
  2FGLJ0017.6-0510 & J001735.81-051241.6 & BZQJ0017-0512 & 1.06(0.03)& 2.73(0.04) & 2.45(0.08) & 0.227   & UND & B & 0\\
\end{longtable}
\end{center}

\newpage
\begin{table*}
	\tiny
	\begin{center}
    	\begin{threeparttable}
    	\caption{List of all candidate blazars associated to the 2FB sources. The explanation of the content of the columns can 
	be found in the notes of Table~\ref{tab:catalogsample2fb}.}
	\begin{tabular}{lllccccclclc}
    	\tableline\tableline
	2FGL\tnote{a}  &  \wse\tnote{b}   &  other\tnote{c}     & [3.4]-[4.6]\tnote{d}     & [4.6]-[12]\tnote{e}     & [12]-[22]\tnote{f}     & type\tnote{g}     & class\tnote{h}& notes\tnote{i} & z\tnote{l}    & reassoc.\tnote{m}  & $N_{\mathrm{BR}}$\tnote{n}\\
	name                 &  name                 &  name              	   &     mag             &    mag             &   mag                 &                  &               &               &               & flag               &                         \\ 
	 \tableline    
  2FGLJ0303.5+4713 &    J030335.23+471616.3 &    BZBJ0303+4716           &    1.06(0.03) &     2.88(0.03)  &     2.31(0.03)  &     UND   &     A     &     M     &     ?      &     y     &     0    \\
  2FGLJ0719.3+3306 &    J071919.41+330709.7 &    BZUJ0719+3307           &    1.04(0.03) &     2.71(0.03)  &     2.23(0.03)  &     UND   &     A     &     M     &     0.779  &     y     &     0 \\   
  2FGLJ0730.2-1141 &    J073019.11-114112.5 &    BZQJ0730-1141           &    1.04(0.03) &     2.95(0.03)  &     2.44(0.04)  &     BZQ   &     A     &     M     &     1.589  &     y     &     0   \\ 
  2FGLJ0825.9-2229 &    J082601.57-223027.2 &    BZBJ0826-2230           &    0.96(0.03) &     2.7 (0.03)  &     2.2 (0.04)  &     UND   &     A     &     M     &     0.91   &     y     &     0    \\
  2FGLJ1058.4+0133 &    J105829.61+013358.8 &    BZUJ1058+0133           &    1.1 (0.03) &     2.92(0.03)  &     2.39(0.03)  &     BZQ   &     A     &     M     &     0.89   &     y     &     0    \\
  2FGLJ1153.2+4935 &    J115324.46+493108.9 &    BZQJ1153+4931           &    1.1 (0.03) &     2.8 (0.03)  &     2.34(0.03)  &     UND   &     A     &     M,v   &     0.334  &     y     &     0    \\
\end{tabular}
	\end{threeparttable}
	\end{center}
\end{table*}

\end{document}